\documentclass{siamonline1116}

\usepackage{braket,amsfonts}

\usepackage{array}

\usepackage[caption=false]{subfig}
\usepackage{pgfplots}

\newsiamthm{claim}{Claim}
\newsiamremark{rem}{Remark}
\newsiamremark{expl}{Example}
\newsiamremark{hypothesis}{Hypothesis}
\crefname{hypothesis}{Hypothesis}{Hypotheses}

\usepackage{algorithmic}

\usepackage{graphicx,epstopdf}

\Crefname{ALC@unique}{Line}{Lines}

\usepackage{amsopn}

\numberwithin{theorem}{section}

\usepackage{xspace}
\usepackage{bold-extra}
\usepackage[most]{tcolorbox}

\colorlet{texcscolor}{blue!50!black}
\colorlet{texemcolor}{red!70!black}
\colorlet{texpreamble}{red!70!black}
\colorlet{codebackground}{black!25!white!25}

\usepackage{amsmath,amssymb, amsfonts}
\usepackage{setspace}
\usepackage[mathscr]{euscript}
\usepackage{graphicx}
\usepackage{color}
\usepackage{cite}
\usepackage{geometry}
\usepackage{booktabs} 
\usepackage{enumerate}

\newcommand{\R}{\mathbb R}

\newcommand{\LL}{\mathscr L}
\newcommand{\G}{\mathscr G}

\newcommand{\dm}{\delta_{\text{min}}}
\newcommand{\Lm}{L_{\text{min}}}
\newcommand{\C}{\operatorname{Cov}}

 \newcommand{\dd}{{\mathrm{d}}}
\newcommand{\YObs}{Y_{{\rm Obs}}}
\newcommand{\Ysim}{Y_{{\rm sim}}}
\newcommand{\Ynwp}{Y_{{\rm NWP}}}

\newcommand{\LLnwp}{\mathscr{L}_{\rm NWP}}
\newcommand{\LLobs}{\mathscr{L}_{\rm Obs \vert NWP}}

\lstdefinestyle{siamlatex}{%
  style=tcblatex,
  texcsstyle=*\color{texcscolor},
  texcsstyle=[2]\color{texemcolor},
  keywordstyle=[2]\color{texemcolor},
  moretexcs={cref,Cref,maketitle,mathcal,text,headers,email,url},
}

\tcbset{%
  colframe=black!75!white!75,
  coltitle=white,
  colback=codebackground, % bottom/left side
  colbacklower=white, % top/right side
  fonttitle=\bfseries,
  arc=0pt,outer arc=0pt,
  top=1pt,bottom=1pt,left=1mm,right=1mm,middle=1mm,boxsep=1mm,
  leftrule=0.3mm,rightrule=0.3mm,toprule=0.3mm,bottomrule=0.3mm,
  listing options={style=siamlatex}
}

\newtcblisting[use counter=example]{example}[2][]{%
  title={Example~\thetcbcounter: #2},#1}

\newtcbinputlisting[use counter=example]{\examplefile}[3][]{%
  title={Example~\thetcbcounter: #2},listing file={#3},#1}

\DeclareTotalTCBox{\code}{ v O{} }
{ %fontupper=\ttfamily\color{texemcolor},
  fontupper=\ttfamily\color{black},
  nobeforeafter,
  tcbox raise base,
  colback=codebackground,colframe=white,
  top=0pt,bottom=0pt,left=0mm,right=0mm,
  leftrule=0pt,rightrule=0pt,toprule=0mm,bottomrule=0mm,
  boxsep=0.5mm,
  #2}{#1}

\patchcmd\newpage{\vfil}{}{}{}
\begin{tcbverbatimwrite}{tmp_\jobname_header.tex}
\title{Global sensitivity analysis for statistical model parameters
  \thanks{This material was based upon work partially supported by the National
Science Foundation under Grant DMS-1127914 to the Statistical and
Applied Mathematical Sciences Institute. Any opinions, findings, and
conclusions or recommendations expressed in this material are those of
the authors and do not necessarily reflect the views of the National
Science Foundation. This material is also partially based upon work
supported by the U.S.\ Department of Energy, Office of Science,
Advanced Scientific Computing Research Program under contract
DE-AC02-06CH11357 (FWP \#57820). }}

\author{Joseph Hart
  \thanks{Department of Mathematics, North Carolina State University, Raleigh, NC (\email{jlhart3@ncsu.edu}).}
  \and
  Julie Bessac
   \thanks{Mathematics and Computer Science Division, Argonne National Laboratory, Lemont, IL (\email{jbessac@anl.gov }).}
  \and
  Emil Constantinescu
  \thanks{Mathematics and Computer Science Division, Argonne National Laboratory, Lemont, IL, The University of Chicago, Chicago, IL, (\email{emconsta@mcs.anl.gov}).}
}

\headers{Global sensitivity analysis for statistical model parameters}
{Joseph Hart, Julie Bessac, and Emil Constantinescu}
\end{tcbverbatimwrite}
\input{tmp_\jobname_header.tex}

\ifpdf
\hypersetup{ pdftitle={Guide to Using  SIAM'S \LaTeX\ Style} }
\fi

\begin{document}
\maketitle

%% ------------------------------------------------------------------
%% ABSTRACT
%% ------------------------------------------------------------------
\begin{tcbverbatimwrite}{tmp_\jobname_abstract.tex}
\begin{abstract}
Global sensitivity analysis (GSA) is frequently used to analyze the
influence of uncertain parameters in mathematical models and simulations. In
principle, tools from GSA may be extended to analyze the influence of
parameters in statistical models. Such analyses may enable reduced or 
parsimonious modeling and greater predictive capability. However,
difficulties such as parameter correlation, model stochasticity,
multivariate model output, and unknown parameter distributions prohibit a direct application of GSA tools
to 
statistical models. 
By leveraging a loss function associated with the statistical model, we introduce a novel framework to address
these difficulties and enable efficient GSA for statistical model
parameters. 
Theoretical and computational properties are considered
and illustrated on a synthetic example. 
The framework is applied
to a Gaussian process model from the literature, which depends on 95 parameters.
Non-influential parameters are discovered through GSA and
a reduced model with equal or stronger predictive capability 
is constructed by using only 79 parameters.
\end{abstract}


\end{tcbverbatimwrite}
\input{tmp_\jobname_abstract.tex}
%% ------------------------------------------------------------------
%% END HEADER
%% ------------------------------------------------------------------

\section{Introduction}

Global sensitivity analysis (GSA) aims to quantify the
relative importance of input variables or factors in determining the
value of a function \cite{saltellibook, intro_SA_uq_handbook}.
It has been used widely for analysis of parameter uncertainty in
mathematical models and simulations \cite{saltellibook, iooss}. In particular, GSA
may be used to improve modeling insight, encourage model parsimony,
and accelerate the model-fitting process. In this paper we propose a novel method for GSA of parameters in statistical models.
To our knowledge, the GSA
tools developed for mathematical models and simulations have not been systematically
developed for analysis of statistical models. The
combination of parameter correlation, model stochasticity, 
multivariate model output, and having an unknown parameter distribution prohibits a direct application of GSA tools to
statistical models. Nevertheless, problem structure in statistical models
may be exploited to enable efficient GSA. 
This paper provides a framework to use existing GSA tools along with tools from statistics to address these challenges and yield a new GSA approach for analysis of statistical models.

This work is motivated by a statistical model that fuses two datasets of
atmospheric wind speed in order to provide statistical prediction of wind speed in space and time \cite{bessac16}.  
The predictions are generated from a Gaussian process whose mean and covariance are parameterized through a large number of parameters, which are determined by numerical optimization. Because of changing weather patterns, the parameters must be re-optimized on a regular basis. Our objective in this application is to reduce the dimension of the parameter space making the model easier to fit and interpret. The optimization procedure to fit parameters is an important problem feature influencing our approach to GSA.
Our method is developed in an abstract setting and subsequently used to analyze the Gaussian process wind speed model.

There are a variety of methods under the umbrella of GSA, the most common is variance-based \cite{sobol,sobol93,saltellialgorithm,variance_based_uq_handbook}. For reasons of computational efficiency, derivative-based methods \cite{lamboni,dgsm1,dgsm2,dgsm_uq_handbook,dgsm_poincare,dgsm_3} have also gained attention recently; they are related to the classical Morris method \cite{morris}. Theoretical challenges in the aforementioned methods has motivated interest in alternatives such as moment-independent importance measures \cite{miim_uq_handbook, borgonovo1, borgonovo2}, Shapley effects \cite{owen_prieur_shapley,staum,iooss_prieur_shapley,owen_shapley_sobol}, and dependence measures \cite{dependence_measures}. In this paper we propose a derivative-based strategy for GSA of statistical models. In principle, any of the aforementioned methods may be used. Computational considerations make derivative-based methods preferable. In particular, the number of gradient evaluations is independent of the parameter space dimension (evaluating the gradient may depend on the dimension but can frequently be evaluated efficiently) and they do not require sampling from conditional distributions.

To perform GSA, a probability distribution must be defined on the input space, and the analysis is done with respect to it. 
GSA has been well developed and studied for problems where the inputs are independent. 
In many statistical models, however, the inputs (parameters) are correlated, thus posing additional challenges to traditional GSA tools. Developing GSA tools for problems with correlated inputs is an active area of research \cite{xugertner,xufast,li,mara,chastaing,chastaingalgorithm,borgonovo1,zhou,borgonovo2,staum,miim_uq_handbook,owen_shapley_sobol,owen_prieur_shapley,iooss_prieur_shapley}. In addition to the theoretical and computational challenges posed by input correlations, simply defining or fitting a joint distribution on the inputs may be challenging in the context of statistical models.

The traditional GSA framework has focused on real-valued deterministic functions $\G:\R^n \to \R$ with uncertain inputs.
A space-time Gaussian process, as in our motivating application, is not a deterministic real-valued function but rather a stochastic vector-valued function, i.e. $\G:\R^n \to S$, where $S$ is a set of random vectors.
Real-valued stochastic processes are considered in \cite{hart,marrel3} and vector-valued deterministic models in \cite{vectoroutput,spatiotemporal_uq_handbook}; generalizing GSA tools to stochastic and/or vector-valued functions is an area of ongoing research. In principle, these approaches may be used to compute sensitivities of a statistical model with respect to its parameters; however, treating them as generic stochastic processes fails to exploit important structure in statistical models.
Sensitivity analysis of statistical surrogate models is considered in \cite{oakley2004,kennedy2006}; however, they focus on the sensitivity of model inputs instead of model parameters. The work of \cite{GP_sensitivity_priors} considers the sensitivity of a Gaussian process model to changes in the prior and correlation function. To the author's knowledge, the approach proposed in this article is the first to formally apply methods from the global sensitivity analysis literature to analyze the influence of statistical model parameters.

 This article provides a framework to
connect the mathematical and statistical tools needed to efficiently extend GSA to
statistical models. We use the loss function associated with the statistical model parameter estimation to define a joint probability distribution, which respects the correlation structure in the problem. 
This distribution is sampled using a Markov Chain Monte Carlo method, and derivative-based sensitivity indices of the loss function are computed from these samples. In this framework, we are able to discover both sensitivities and correlation structures without requiring a priori knowledge about the parameters.

Our framework requires efficient evaluations of the loss function's gradient. In fact, it is designed to exploit efficient gradient evaluations, as is the case in our motivating application. Derivative-based methods are typically limited to identifying unimportant parameters as they may fail to capture the relative importance of the most influential parameters. Though generally undesirable, this is permissible when seeking model simplification and/or parameter dimension reduction as identifying unimportant parameters is the primary focus.

We provide definitions and construct the abstract framework in Section~\ref{definitions}. Section~\ref{methodology} details the computational methodology and summarizes the proposed method. In Section~\ref{results}, we present numerical results for a synthetic test problem and for our motivating application. 
Section \ref{conclusion} provides a brief summary of our conclusions. 

\section{Preliminaries}
\label{definitions}

Let $\G$ be a statistical model defined through parameters\\
$\theta=(\theta_1,\theta_2,\dots,\theta_n)^T \in A$, $A \subseteq
\R^n$. Let $\LL(.;y):A \to \R$ be a loss function (or cost function) associated with $\G$;
that is, to fit $\G$ to the data $y$, one computes $\arg\min_{\theta \in A}
\LL(\theta;y)$.  
In the model-oriented context of statistics, parameters of the model are usually estimated by minimizing $\LL(\cdot;y)$ computed on observed data $y$. 
The loss function is chosen given the model and its intended use. Common examples are least-squares and maximum likelihood. For simplicity, $\LL$ will be used instead of $\LL(\cdot;y)$ for the remainder of the paper.
In our motivating application, $\G$ is a Gaussian process with
mean $\mu(\theta)$ and covariance $\Sigma(\theta)$ depending on a vector of
parameters $\theta$ and $\LL$ is the the negative log likelihood. We assume that $\nabla \LL$ may be computed efficiently, as is the case in our motivating application.

We are interested in the global
sensitivity of $\G$ with respect to $\theta$.
Since $\G$ may be a complex mathematical object, we propose to analyze the global
sensitivity of $\G$ to $\theta$ through the global sensitivity of
$\LL$ to $\theta$. 
This makes our analysis dependent on the choice of loss function, which is consistent with a goal oriented choice of $\LL$. This is appropriate since $\LL$ encodes the
dependence of $\G$ on $\theta$. Further, since $\LL$ is a
deterministic real-valued function of $\theta$, it is frequently
easier to analyze than $\G$.

One challenge is that the statistical model $\G$ may be mathematically
well defined for parameters $\theta \in A$ but 
yield a practically irrelevant solution in the context of a given
application. To avoid this scenario, we let $B \subseteq A$ be the
subset that restricts $A$ to 
parameters yielding relevant solutions. For
instance, a quantity in $\G$ may be required to be nonnegative so $B$
restricts to parameters respecting this constraint. We assume that $B$
is a Lebesgue measurable set; this is easily verified in most applications.

We make three assumptions about $\LL$; formally, they are expressed as the following.

\begin{enumerate}[I]
\item $\LL$ is differentiable. \label{a1}
\item $\exists$ $\dm \ge 0$ such that $\int_B e^{-\dm \LL(\theta)}
  \dd \theta$, $\int_B \vert \theta_k \vert e^{-\dm \LL(\theta)} \dd \theta$, and\\ $\int_B \vert \frac{ \partial \LL}{\partial \theta_k}
  (\theta) \vert e^{-\dm \LL(\theta)} \dd \theta$, $k=1,2,\dots,n$, exist
  and are finite. \label{a2}
\item $\exists$ $\Lm \in \R$ such that $\Lm \le \LL(\theta)$, $\forall \theta \in B$. \label{a3}
\end{enumerate}

Assumption~\ref{a1} is necessary since we seek to use a
derivative-based GSA. This assumption is easily verifiable in most cases. Assumption~\ref{a2} is needed so that global sensitivity indices \eqref{S} are well defined. Assumption~\ref{a3} is a needed technical assumption requiring that the loss function be bounded below. Note that if $\LL$ is continuously differentiable and $B$ is compact, then all three assumptions follow immediately; this is a common case.

To define global sensitivity indices, we must specify a probability measure to integrate against. Let
\begin{eqnarray}
q(\theta)=\chi_B(\theta)e^{- \delta ( \LL(\theta)+\lambda \vert \vert \theta \vert \vert_2^2 )}
\label{q}
\end{eqnarray}
for some $\delta \ge \dm$ and $\lambda \ge 0$; $\chi$ is the characteristic function of a
set, and $\vert \vert \cdot \vert \vert_2$ is the Euclidean norm. Note that $B$ is defined through constraints on $\G$ so it is generally difficult to express $B$ in terms of simple algebraic constraints. 
In most cases, however, the constraints may be checked when $\LL$ is evaluated, and hence $q$ is easily evaluated through evaluating $\LL$.

From Assumption~\ref{a2} and the fact that $e^{-\delta \lambda \vert \vert \theta \vert \vert_2^2} \le 1$, it follows that $q$ is integrable. 
We define the probability density function (PDF) as
\begin{eqnarray}
p(\theta)=\frac{q(\theta)}{\int_A q(\gamma) \dd \gamma}=\frac{e^{- \delta ( \LL(\theta)+\lambda \vert \vert \theta \vert \vert_2^2 )}}{\int_B e^{- \delta ( \LL(\gamma)+\lambda \vert \vert \gamma \vert \vert_2^2 )} \dd \gamma} .
\label{p}
\end{eqnarray}

Then $p$ is supported on $B$ and gives the greatest probability to regions where $\LL$ is close to its minimum namely, where $\theta$ is a good fit. A PDF of this form corresponds to a Gibbs measure \cite{gibbs} with temperature $\delta$; the temperature determines how the probability mass disperses from the modes. The scalar $\lambda \ge 0$ is a regularization factor that aids when $p$ is too heavy tailed; this is illustrated in Section~\ref{results}. The determination of $\delta$ and $\lambda$ is considered in Section~\ref{methodology}. Our formulation shares similar characteristics to Bayesian inference. For instance, if $\LL$ is a negative log likelihood and $\delta=1$ then $\eqref{p}$ is the posterior PDF of $\theta$ using a Gaussian prior truncated to $B$ (or when $\lambda=0$, the prior is simply a uniform distribution on $B$). 
%  Note that determining a PDF for correlated parameters may be challenging in general. This framework provides a natural way to define the PDF, but it comes at the cost of needing to determine $\delta$ and $\lambda$.
%
\begin{definition} Let the sensitivity index of $\G$ with respect
  to $\theta_k$ be defined as
\begin{eqnarray}
S_k=\mathbb E \left( \vert \theta_k \vert \right) \mathbb E \left( \left\vert \frac{\partial \LL}{\partial \theta_k} (\theta) \right\vert \right)=\int_B \vert \theta_k \vert p(\theta)\dd \theta  \int_B \left\vert \frac{\partial \LL}{\partial \theta_k} (\theta) \right\vert p(\theta)\dd \theta .
\label{S}
\end{eqnarray}
\end{definition}
Derivative-based sensitivity indices are commonly defined in the literature by taking the expected value of the partial derivative squared. The absolute value is used here because $\LL$ and its derivatives typically become large for parameters with low probability, so squaring the partial derivative results in the low probability realizations making larger contributions to \eqref{S}. Since $\mathbb E \left( \left\vert \frac{\partial \LL}{\partial \theta_k} (\theta) \right\vert \right)$ depends on the units of $\theta_k$, it will be difficult to compare partial derivatives when parameters are on multiple scales. Typically one would rescale parameters a priori to avoid this issue, but this is difficult to do in our context. Multiplying by $\mathbb E\left( \vert \theta_k \vert \right)$ yields scale-invariant global sensitivity indices. Sensitivity indices for groups of variables \cite{dgsm2} may also be defined in our framework, but are not considered in this paper.

Correlations in $\theta$ make Monte Carlo integration with uniform
sampling intractable for computing the $S_k$'s. Importance sampling
may be used if an efficient proposal distribution is found; however,
this is also challenging in most cases. Therefore, we propose to compute the $S_k$'s with Markov Chain Monte Carlo (MCMC) methods.

In summary, the global sensitivity of $\G$ to $\theta$ may be
estimated by using only evaluations of $\LL$ and $\nabla \LL$ along with
MCMC. This framework also admits additional useful information as
by-products of estimating \eqref{S}. More details are given in Section~\ref{methodology}.

\section{Computing sensitivities\label{methodology}}
In this section we present the main result of this study. The proposed method may be partitioned into three stages:
\begin{enumerate}[i]
\item Preprocessing where we collect information about the loss function, \label{s1}
\item Sampling where samples are drawn from the probability measure \eqref{p}, \label{s2}
\item Post-processing where sensitivities as well as additional information are computed. \label{s3}
\end{enumerate}
In the preprocessing stage we seek to find characteristic values for the parameters and the loss function. These characteristic values are used to determine the temperature and regularization factor in the PDF $p$ \eqref{p}.

In the sampling stage we first determine the temperature and regularization factor. Subsequently an MCMC sampler is run to collect samples from \eqref{p}.

In the post-processing stage we compute sensitivities by evaluating the gradient of the loss function at the samples drawn in the sampling stage \ref{s2}. In addition, the robustness of the sensitivities with respect to perturbations in the temperature and the parameter correlations are extracted from the existing samples and gradient evaluations. These two pieces of information are by-products of computing sensitivities and require no additional computation.

These three stages are described in
Subsections~\ref{stage_1},~\ref{stage_2}, and ~\ref{stage_3}, respectively. The method is summarized as a whole in Subsection~\ref{summary}.

\subsection{Preprocessing stage}
\label{stage_1}
Characteristic magnitudes for $\theta$ and $\LL$ are needed to determine the regularization factor and temperature. To this end we introduce two auxiliary computations as a preprocessing step.

The first auxiliary computation runs an optimization routine to
minimize $\LL$; the choice of optimizer is not essential here. Let
$\theta^{\star}$ be the minimizing parameter vector. For our purposes
it is acceptable if $\theta^{\star}$ is not the global minimizer of
$\LL$ as long as it is sufficiently close to capture characteristic
magnitudes of $\LL$ in regions of good fit.

The second auxiliary computation uses $\theta^{\star}$ to determine the range of loss function values that our MCMC sampler should explore. Let $c>0$, and let $\Theta=(\Theta_1,\Theta_2,\dots,\Theta_n)$ be a random vector defined by
\begin{eqnarray}
\label{theta_unif}
\Theta_k \sim \mathcal U[(1-c)\theta_k^\star,(1+c) \theta_k^\star],
\end{eqnarray}
where all the $\Theta_k$'s are independent of one another and $\mathcal U$ denotes the uniform distribution. Hence, $\Theta$ represents uniform uncertainty of $c \%$ about $\theta^\star$.

Determining $c$ is an application-dependent problem. 
In fact, its determination is the only portion of our proposed method that cannot be automated. 
To choose $c$, we suggest fixing a value for $c$, sampling from $\Theta$, and assessing the quality of $\G$'s predictions using the sample. Repeating this sample and assessment process for various values of $c$ allow the user to determine a $c$ that yields reasonable predictions. This step is highly subjective and application dependent; however, it is a very natural means of inserting user specification. One simple way to do this is visualizing the model output for each sample and increasing $c$ until the outputs become unrealistic.

Taking large values for $c$ will result in the PDF $p$ giving significant probability to regions of the parameter space yielding poor fits, and thus hence sensitivity indices that are not useful. Taking small values for $c$ will result in the PDF $p$ giving significant probability to regions of the parameter space near local minima, thus making the sensitivity indices local. Since the choice of $c$ is strongly user dependent, the robustness of the sensitivity indices with respect to perturbations in $c$ is highly relevant; this is indirectly addressed by Theorem~\ref{diffthm}.

Once $c$ is specified, then a threshold $M$, which is used to compute the regularization factor and temperature (see Subsection~\ref{regular} and Subsection~\ref{delta}), may be easily computed via Monte Carlo integration. 
We define the threshold
\begin{eqnarray}
M=\mathbb E(\LL(\Theta)).
\label{M}
\end{eqnarray}
Note that the expectation in \eqref{M} is computed with respect to the independent uniform measure; all other expectations in the paper are computed with respect to the PDF $p$ \eqref{p}. 

\subsection{Sampling stage \label{stage_2}}
We use an MCMC method to sample from $p$ \eqref{p} through evaluations of the unnormalized density $q$ \eqref{q}. Then the $S_k$'s may be computed through evaluations of $\nabla \LL$ at the sample points. Many MCMC methods may be used to sample $p$; see, for example \cite{sherlock, rosenthal, vihola,geyer,geyerbook,dram}.

Determining which MCMC method to use and when it has converged may be challenging. 
Convergence diagnostics \cite{convdiag,diagnostic_review} have been developed that may identify when the chain has not converged; however, they all have limitations and cannot ensure convergence \cite{intromcmc}. 
In Section~\ref{results}, adaptive MCMC \cite{vihola} is used with the convergence diagnostic from \cite{mpsrf}.

Assuming that an MCMC sampler is specified, we focus on determining the temperature and regularization factors in Subsections \ref{regular} and \ref{delta}, respectively.

\subsubsection{Determining the regularization factor}
\label{regular}
To determine the regularization factor $\lambda$, consider the function
\begin{eqnarray*}
\LL_\lambda(\theta)=\LL(\theta)+\lambda \vert \vert \theta \vert \vert_2^2 .
\end{eqnarray*}
The PDF $p$ gives greatest probability to regions where $\LL_\lambda$ is small. If $\lambda \vert \vert \theta \vert \vert_2^2$ is small relative to $\LL(\theta)$, then the local minima of $\LL_\lambda$ are near the local minima of $\LL$. Ideally we would like $\lambda=0$, but in some cases this results in $p$ being too heavy tailed. Instead we may require that $\lambda \vert \vert \theta \vert \vert_2^2 \approx \nu \LL_\lambda(\theta)$ for some $\nu \in (0,1)$; that is, the regularization term contributes $\nu$ percent of the value of $\LL_\lambda$. Setting $\lambda \vert \vert \theta \vert \vert_2^2 = \nu \LL_\lambda(\theta)$ and replacing $\LL(\theta)$ and $\theta$ with $M$ and $\theta^{\star}$, we get
\begin{eqnarray}
\lambda=\frac{\nu M}{(1-\nu) \vert \vert \theta^{\star} \vert \vert_2^2}.
\label{lambda}
\end{eqnarray}
In practice we suggest beginning with $\nu=0$. 
If the MCMC sampler yields heavy-tailed distributions that converge slowly, then $\nu$ may be increased to aid the convergence. 
This case is illustrated in Section~\ref{results}.

\subsubsection{Determining the temperature}
\label{delta}
To determine the temperature $\delta$, we first define
\begin{eqnarray*}
M_\lambda=M+\lambda \vert \vert \theta^{\star} \vert \vert_2^2 .
\end{eqnarray*}
 We seek to find $\delta$ so that  $\LL_\lambda(\theta) \le M_\lambda$ with probability $\alpha$; $\alpha=.99$ is suggested to mitigate wasted computation in regions where $\theta$ yields a poor fit. Let $C=\{ \theta \in B \vert \LL_\lambda(\theta) \le M_\lambda \}$.
We note that $C$ is a Lebesgue measurable set since $\LL_\lambda$ is continuous and $B$ is Lebesgue measurable. 
We define the function $\Delta:[\dm, \infty) \to [0,1]$ by 
\begin{eqnarray}
\label{delta_def_eq}
\Delta(\delta)=\int_C p(\theta)\dd \theta.
\end{eqnarray}
Then $\Delta(\delta)$ gives the probability that $\LL_\lambda(\theta)
\le M_\lambda$. The optimal temperature $\delta$ is the solution of
$\Delta(\delta)=\alpha$. Four results are given below showing that
$\Delta$ possesses advantageous properties making the nonlinear equation
$\Delta(\delta)=\alpha$ easily solvable. The proofs of the following
propositions are given in the
appendix. 
\begin{proposition}
\label{Deltadiff}
If $\int_B \LL_\lambda(\theta) e^{-\dm \LL_\lambda(\theta) } \dd \theta<\infty$, then $\Delta$ is differentiable on $(\dm,\infty)$ with 
\begin{eqnarray}
\Delta'(\delta)=(-1+\Delta(\delta)) \int_C \LL_\lambda(\theta) p(\theta) \dd \theta + \Delta(\delta) \int_{B \setminus C} \LL_\lambda(\theta) p(\theta) \dd \theta .
\end{eqnarray}
\end{proposition}

\begin{proposition}
\label{incfun}
If $\int_B \LL_\lambda(\theta) e^{-\dm \LL_\lambda(\theta) } \dd \theta<\infty$, then $\Delta$ is a strictly increasing function on $(\dm,\infty)$.
\end{proposition}

Propositions~\ref{Deltadiff} and~\ref{incfun} yield desirable properties of $\Delta$. The assumption that\\ $\LL_\lambda(\theta)e^{-\dm \LL_\lambda(\theta)}$ is integrable is necessary for $\Delta'(\theta)$ to be well defined. Note that this assumption follows from Assumption~\ref{a1} when $B$ is bounded. Theorem~\ref{deltalimit} and Corollary~\ref{existence} below give existence and uniqueness, respectively, for the solution of $\Delta(\delta)=\alpha$ under mild assumptions.

\begin{theorem}
\label{deltalimit}
If $B$ is a bounded set and $\exists \theta' \in B$ such that $\LL_\lambda(\theta')<M_\lambda$, then $\forall \alpha \in (0,1)$ $\exists \delta > \dm$ such that $\Delta(\delta) > \alpha$.
\end{theorem}

\begin{corollary}
\label{existence}
If $\alpha \in (\Delta(\dm),1)$, $B$ is a bounded set, and $\exists \theta' \in B$ such that\\ $\LL_\lambda(\theta')<M_\lambda$, then $\Delta(\delta)=\alpha$ admits a unique solution.
\end{corollary}

The assumption that $B$ is bounded is reasonable in most applications; $A$ may be unbounded, but $B$ is restricted to relevant solutions that will typically be bounded. The assumption that $\LL_\lambda(\theta')<M_\lambda$ means that $M_\lambda$ is not chosen as the global minimum, which should always hold in practice. The assumption that $\alpha \in (\Delta(\dm),1)$ is necessary for existence. Typically $\Delta(\dm)$ is much less than 1, while $\alpha$ is chosen close to 1. The assumptions Theorem~\ref{deltalimit} and Corollary~\ref{existence} hold in most applications

In summary, under mild assumptions $\Delta(\delta)=\alpha$ is a scalar nonlinear equation admitting a unique solution and $\Delta$ possesses nice properties (monotonicity and differentiability). Further, $\Delta(\delta)$ and $\Delta'(\delta)$ may be approximated simultaneously by running MCMC. The challenge is that evaluating $\Delta(\delta)$ and $\Delta'(\delta)$ in high precision requires running a long MCMC chain. In fact, $\Delta'(\delta)$ is significantly more challenging to evaluate than $\Delta(\delta)$. For this reason we suggest using derivative-free nonlinear solvers which will still be efficient since $\Delta$ is a well-behaved function. In the spirit of inexact Newton methods \cite{kelley}, shorter chains may be run for the early iterations solving $\Delta(\delta)=\alpha$ and the precision increased near the solution. In practice, relatively few evaluations of $\Delta$ are needed because of its properties, shown above.

As previously highlighted, the PDF \eqref{p} corresponds to a Bayesian posterior PDF when $\LL$ is a negative log likelihood and $\delta=1$.  If $\delta<1$, our GSA approach uses a ``flatter" PDF than the Bayesian posterior. Our determination of $\delta$ incorporates information from the parameter optimization procedure to ensure that our sensitivity analysis searches the parameter space in which the optimization routine traverses.

\subsection{Post-processing stage \label{stage_3}}
Having attained samples from $p$ \eqref{p}, the sensitivities \eqref{S} may be estimated by evaluating $\nabla \LL$ at the sample points and forming the Monte Carlo estimator for the expectations in \eqref{S}. In addition to computing these sensitivities, we may extract two other useful pieces of information, namely, the robustness of the sensitivities with respect to perturbations in the temperature and the parameter correlations. These are described in Subsections \ref{delta_sensitivity} and \ref{correlations}, respectively.
\subsubsection{Robustness with respect to the temperature \label{delta_sensitivity}}
As a result of the uncertainty in the determination of $\delta$ (computation of $\theta^\star$, choice of $c$, estimation of $M$, solution of $\Delta(\delta)=\alpha$), we analyze the robustness of the sensitivities with respect to $\delta$. Consider the functions
\begin{eqnarray*}
F_k: (\dm,\infty) &\to& \R, \\
\delta &\mapsto & \left( \int_B \left\vert \theta_k \right\vert \left( \frac{e^{-\delta \LL_\lambda(\theta)}}{\int_B e^{-\delta \LL_\lambda(\tilde{\theta})} d\tilde{\theta}} \right) \dd \theta \right) \left( \int_B \left\vert \frac{\partial \LL}{\partial \theta_k} (\theta) \right\vert \left( \frac{e^{-\delta \LL_\lambda(\theta)}}{\int_B e^{-\delta \LL_\lambda(\tilde{\theta})} d\tilde{\theta}} \right) \dd \theta \right)
\end{eqnarray*}
$k=1,2,\dots,n$; clearly $F_k(\delta)=S_k$. Theorem~\ref{diffthm} gives the derivative of the sensitivity index with respect to the temperature $\delta$, namely, $F_k'(\delta)$. 
\begin{theorem}
\label{diffthm}
If $\int_B \LL_\lambda(\theta)
e^{-\dm \LL_\lambda(\theta)} \dd \theta$, $\int_B \LL_\lambda(\theta) \vert \theta_k \vert e^{-\dm \LL_\lambda(\theta)} \dd \theta$, and\\
 $\int_B  \LL_\lambda(\theta)  \vert \frac{
  \partial \LL}{\partial \theta_k} (\theta) \vert e^{-\dm \LL_\lambda(\theta)}
\dd \theta$ exist and are finite, then $F_k$ is differentiable with
\begin{eqnarray}
\label{F_prime}
F_k'(\delta)=-\C(\vert \theta_k \vert , \LL_\lambda(\theta) ) \mathbb E \left( \left\vert \frac{\partial \LL}{\partial \theta_k}(\theta)
\right\vert  \right) - \mathbb E \left( \vert \theta_k \vert \right) \C \left(\left\vert \frac{\partial \LL}{\partial \theta_k}(\theta)
\right\vert ,\LL_\lambda(\theta) \right) ,
\end{eqnarray}
where $\C(\cdot,\cdot)$ is the covariance operator.
\end{theorem}

Theorem~\ref{diffthm} allows $F_k'(\delta)$ to be computed from the samples and function evaluations used to compute $S_k$. For small $h$, $F_k(\delta+\delta h) \approx F_k(\delta)+h \delta F_k'(\delta)$ so the robustness of $S_k$ may be estimated without any further computational expense. 

Since the magnitude of $S_k$ may depend on $\delta$, it is useful to normalize for each $h$ when assessing robustness. 
We define
\begin{eqnarray}
\label{Fhat}
\hat{F}_k(\delta+\delta h)=\frac{F_k(\delta)+h\delta F_k'(\delta)}{\sum_{j=1}^n (F_j(\delta)+h\delta F_j'(\delta))}, \quad k=1,2,\dots,n,
\end{eqnarray}
which may be plotted for $h \in (-h_{max},h_{max})$ to assess robustness. Since this is only a local estimate we suggest taking $h_{max}=\frac{1}{10}$, reflecting a $10\%$ uncertainty about $\delta$.

The user may interpret $F_k'(\delta)$ as the local sensitivity of \eqref{S} with respect to $\delta$. Because of the several sources of uncertainty in $\delta$, it is desirable to have a global sensitivity of \eqref{S} with respect to $\delta$; however, this would require significantly more computational effort. Nonetheless, locality in $\delta$ does not diminish the value of \eqref{S} as a global sensitivity index, and it provides useful information about \eqref{S} at a negligible computational cost. 

\subsubsection{Extracting parameter correlations}
\label{correlations}
Parameters are typically correlated, and the correlation information is a valuable complement to the sensitivity indices. For instance, if $\G$ is sensitive to two parameters that are highly correlated, then it may be possible to remove one of them from $\G$ since the other may compensate. In addition, the correlations may reveal parameter misspecifications in $\G$. 

The strength and nature of the correlations in $\theta$ are typically not known a priori.
Correlation coefficients may be computed from the MCMC samples and returned as a by-product of computing sensitivity indices. The Pearson correlation coefficient is commonly used to measure correlations from sampled data. Other measures of correlation may be interchanged within our framework as well.

\subsection{Summary of the method\label{summary}}

This subsection summarizes our proposed method. The method is divided into three algorithms, one for each stage described in Section~\ref{methodology}.

Algorithm~\ref{alg:aux} performs the auxiliary computations of Subsection~\ref{stage_1}. 
Note that determining $c$ in line~2 is the only application-specific portion of the proposed method; user discernment is necessary to choose $c$.

 Algorithm~\ref{alg:sampling} requires the user to specify the parameter $\nu$ from Subsection~\ref{regular}, the parameter $\alpha$ from Subsection~\ref{delta}, and the number of MCMC samples $N$. 
 We suggest starting with $\nu=0$ and rerunning Algorithm~\ref{alg:sampling} with a larger $\nu$ if the convergence results indicate that the PDF is heavy tailed. Hence $\nu$ may be viewed as a computed quantity rather than one specified by the user. As mentioned in Subsection~\ref{delta}, we suggest using $\alpha=.99$. It may be considered fixed and the user only needs to change it if they have a specific purpose which requires giving more weight to ``poor" parameter choices. The choice of $N$ may be difficult; however, more samples may be appended after an initial run so $N$ can be adapted without any wasted computation.
 
 Algorithm~\ref{alg:sensitivities} is a simple post-processing of the MCMC samples to compute sensitivity indices, robustness estimates, and parameter correlations. 
 One may also perform convergence diagnostics on the MCMC estimators of $\mathbb E \left( \left\vert \frac{\partial \LL}{\partial \theta_k}(\theta) \right\vert  \right)$, $k=1,2,\dots,n$, along with Algorithm~\ref{alg:sensitivities}.

\begin{algorithm}
\caption{Auxiliary Computation} \label{alg:aux}
\begin{algorithmic}[1]
\STATE compute $\theta^{\star}=\arg\min \LL(\theta)$ via some
optimization routine 
\STATE determine $c>0$ through visualization of model outputs, see \eqref{theta_unif}
\STATE estimate $M$ via Monte Carlo integration, see \eqref{M}
\end{algorithmic} 
\end{algorithm}

\begin{algorithm}
\caption{Sampling} \label{alg:sampling}
\begin{algorithmic}[1]
\STATE \textbf{function }($\nu$, $N$, $\alpha$)
\STATE compute $\lambda$ using \eqref{lambda}	
\STATE solve $\Delta(\delta)=\alpha$, see \eqref{delta_def_eq}
\STATE run MCMC sampler to draw $N$ samples from $p$ \eqref{p} 
\STATE store MCMC samples in a matrix $X$
\STATE test convergence of the sampler
\STATE \textbf{end function}
\end{algorithmic} 
\end{algorithm}

\begin{algorithm}
\caption{Sensitivities, Perturbations, and Correlations} \label{alg:sensitivities}
\begin{algorithmic}[1]
\STATE evaluate $\nabla \LL$ at points in $X$
\STATE estimate $S_k$ \eqref{S}, $k=1,2,\dots,n$
\STATE estimate $F_k'(\delta)$ \eqref{F_prime}, $k=1,2,\dots,n$
\STATE compute empirical correlation matrices from $X$
\end{algorithmic} 
\end{algorithm}

\section{Numerical results}
\label{results}
In this section we apply the proposed method to two problems. The
first is a synthetic test problem meant to illustrate the methodological
details described in Section~\ref{methodology}. The second is our
motivating application where $\G$ is a space-time hierarchical
Gaussian process used for wind speed forecast \cite{bessac16}.

\subsection{Synthetic test problem}
This synthetic problem illustrates the proposed
method of GSA and its properties on a simple
example with least squares estimation. 
%This synthetic problem validates the proposed
%method of GSA and illustrate its properties on a simple
%example with least squares estimation.
%\commJulie{We consider here a least-square estimation procedure as a simplification of the general case. Indeed in the Gaussian case it is equivalent to the maximum likelihood one, and least-square estimates of individual quantities are often used as initial conditions of maximum likelihood. }
We demonstrate the difficulty of MCMC sampling with heavy tailed distributions and how the regularization factor \eqref{lambda} alleviates this problem. 

Mimicking characteristics of our motivating application, we consider a space-time process governed by the function
\begin{eqnarray}
f(x,t)=S(x)T(t),
\label{masterfun}
\end{eqnarray}
where
\begin{eqnarray*}
S(x)= \alpha_0+\alpha_1 x+\alpha_2x^2\,,~\textnormal{and}
\end{eqnarray*}
\begin{eqnarray*}
T(t)=\beta_0+\beta_1 e^{- \gamma t} \cos \left(\frac{2 \pi}{100} t\right)+\beta_2 \sin \left(\frac{2 \pi}{100} t\right) + \beta_3  \frac{1}{1+e^{-.1(t-50)}}
\end{eqnarray*}
with $x \in [0,1]$, $t \in [0,100]$, and
\begin{align}
  \nonumber
\theta&=(\beta_0,\beta_1,\beta_2,\beta_3,\gamma,\alpha_0,\alpha_1,\alpha_2)
\\
\label{eq:synth:true:theta}
&= (2,10,3,.01,.01,1,.01,1).
\end{align}
We draw $15^2$
samples from \eqref{masterfun} on a uniform grid of $[0,1] \times
[0,100]$, which gives data
\begin{eqnarray*}
\{ (x_i,t_i,f(x_i,t_i) ) \}_{i=1}^{225}.
\end{eqnarray*}

A model $\hat{f}$ parameterized in the same form as \eqref{masterfun} is proposed, but the parameters are assumed to be unknown. They are determined by minimizing the least squares loss function
\begin{eqnarray*}
\LL(\theta)=\frac{1}{225} \sum_{i=1}^{225} (f(x_i,t_i)-\hat{f}(x_i,t_i))^2.
\end{eqnarray*}
Least squares estimates are generally used as initial conditions for maximum likelihood optimization. This motivates a least squares formulation in this example as a simplification of the loss function in our motivating application.

Analytic solutions for the sensitivities are intractable;
however, we can validate our results by comparing them with our
knowledge of the true model that generated the data.
In particular, the relative importance of the parameters is clear by
examining \eqref{masterfun} and
\eqref{eq:synth:true:theta}. We expect $\beta_1$ to be the most important parameter and $\beta_3$ and $\alpha_1$ to be the least important parameters.

\begin{figure}[h]
\begin{centering}
\includegraphics[width=.19 \textwidth]{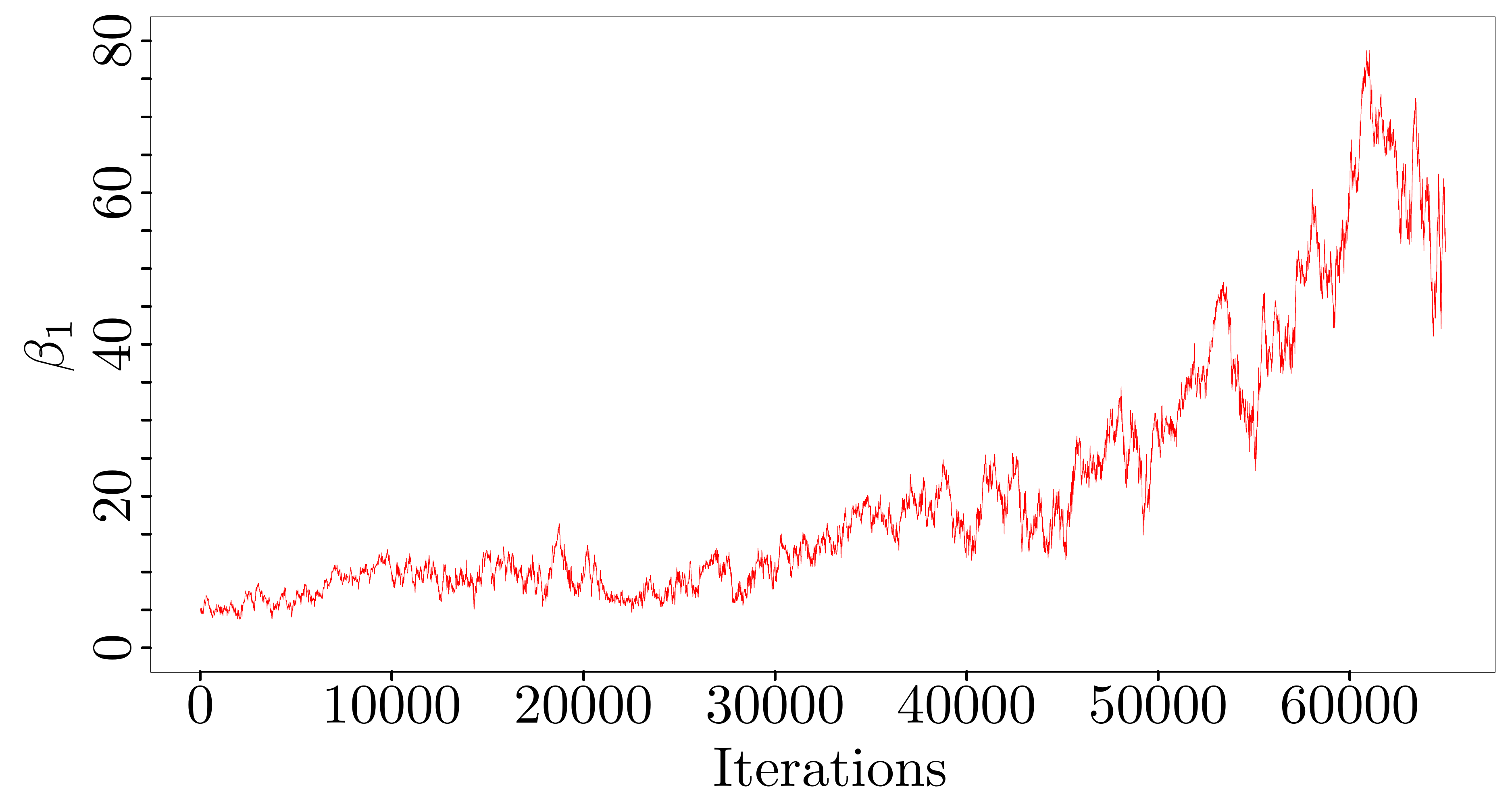}
\includegraphics[width=.19 \textwidth]{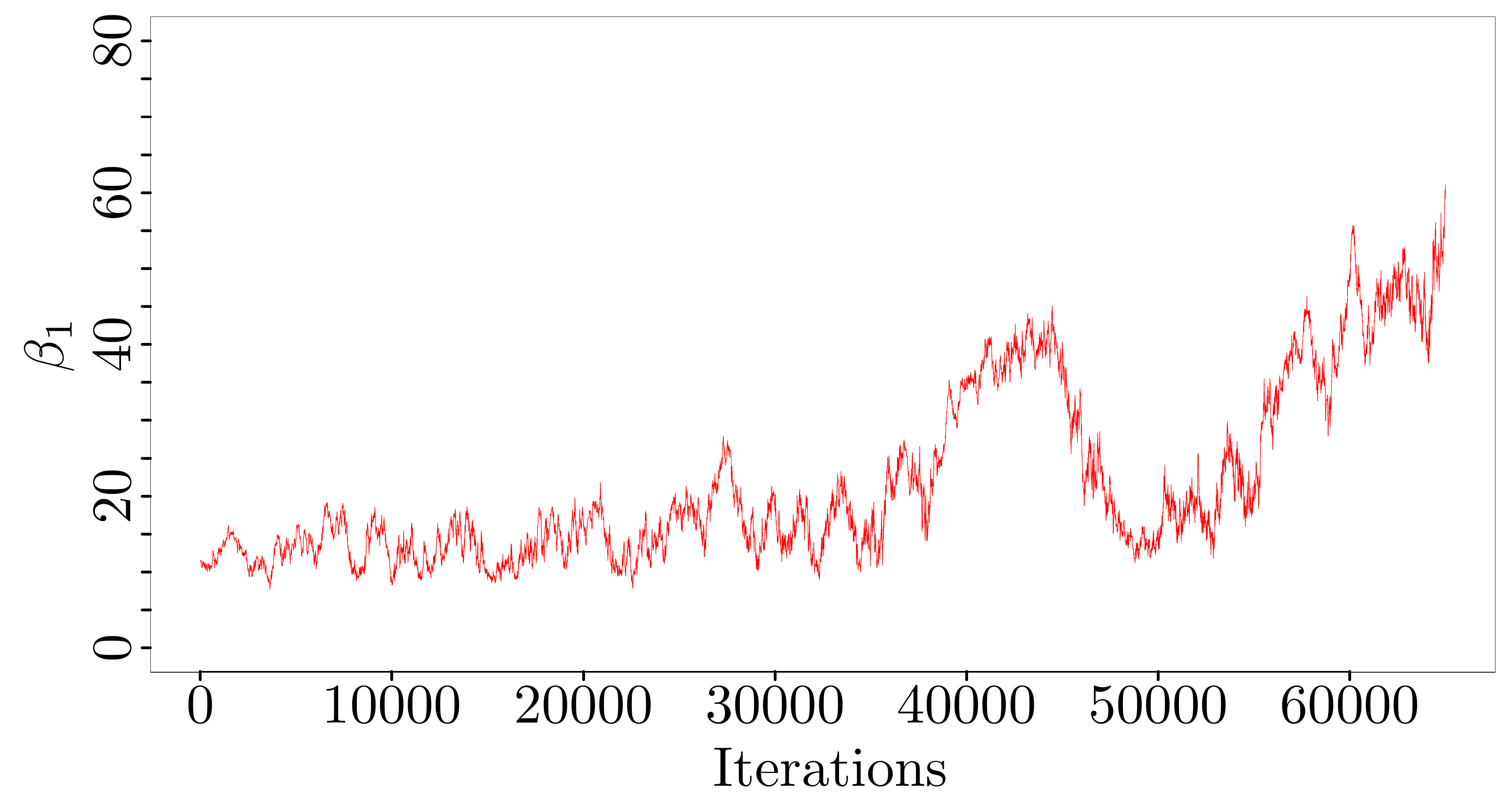}
\includegraphics[width=.19 \textwidth]{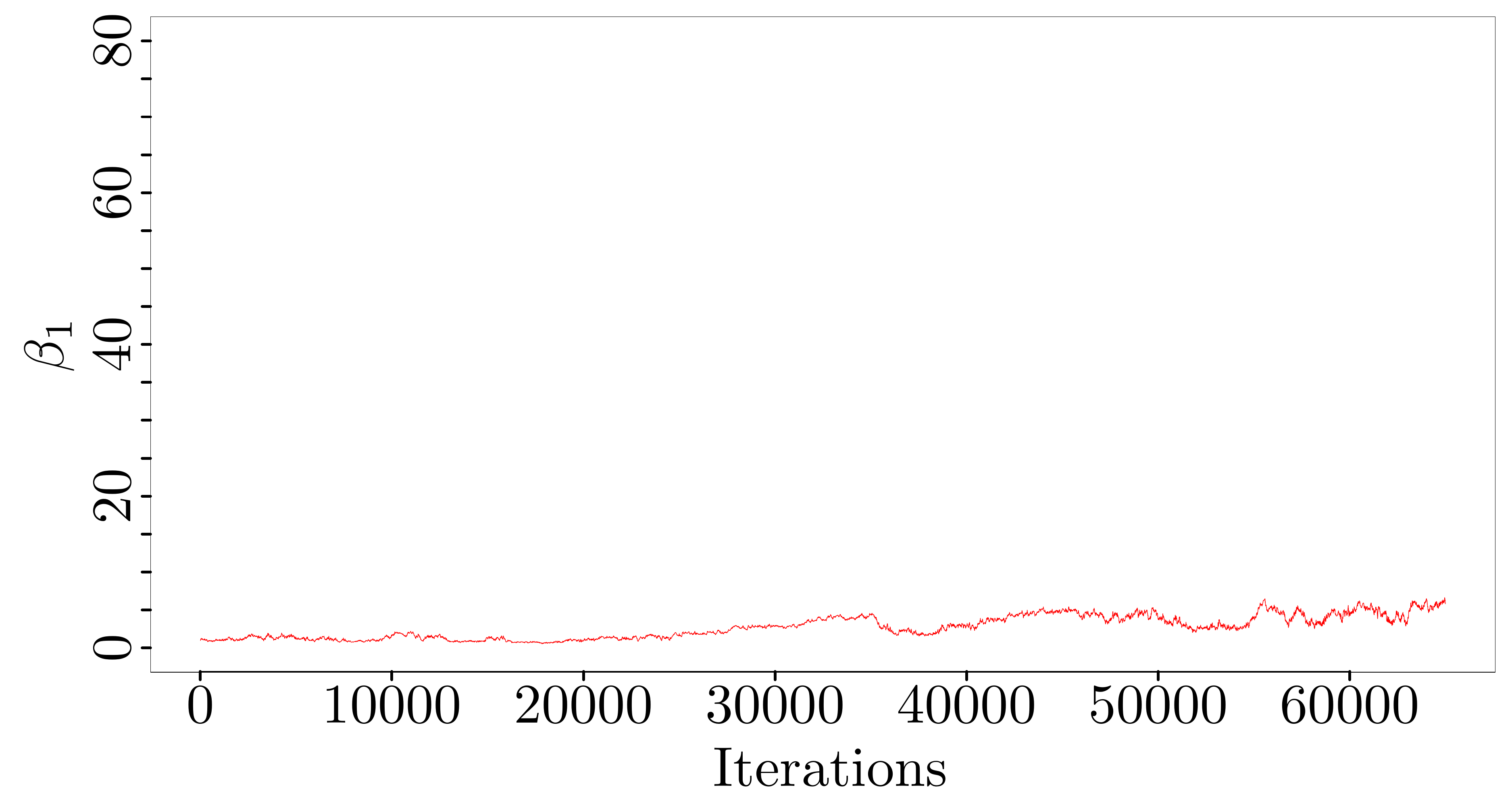}
\includegraphics[width=.19 \textwidth]{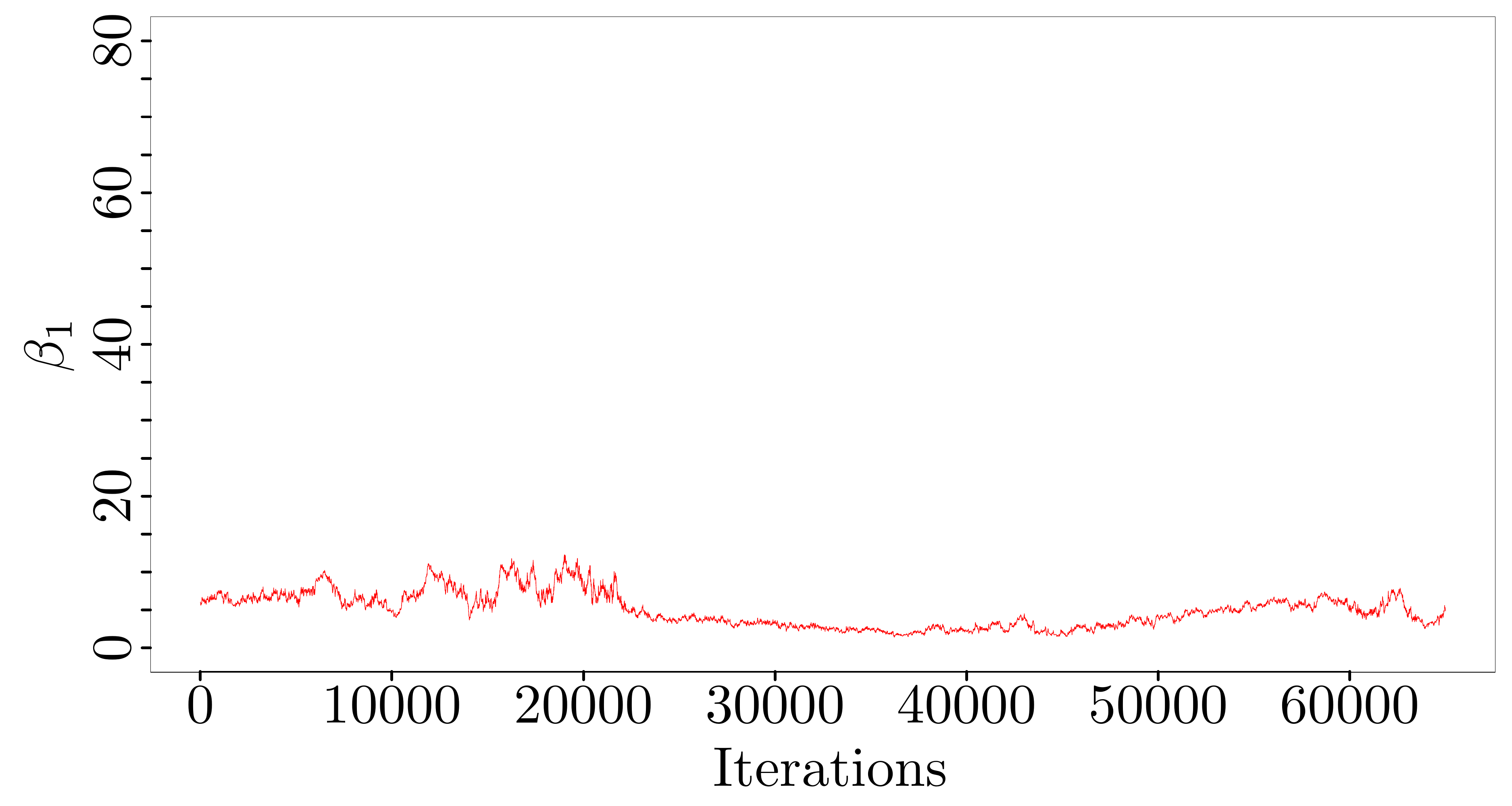}
\includegraphics[width=.19 \textwidth]{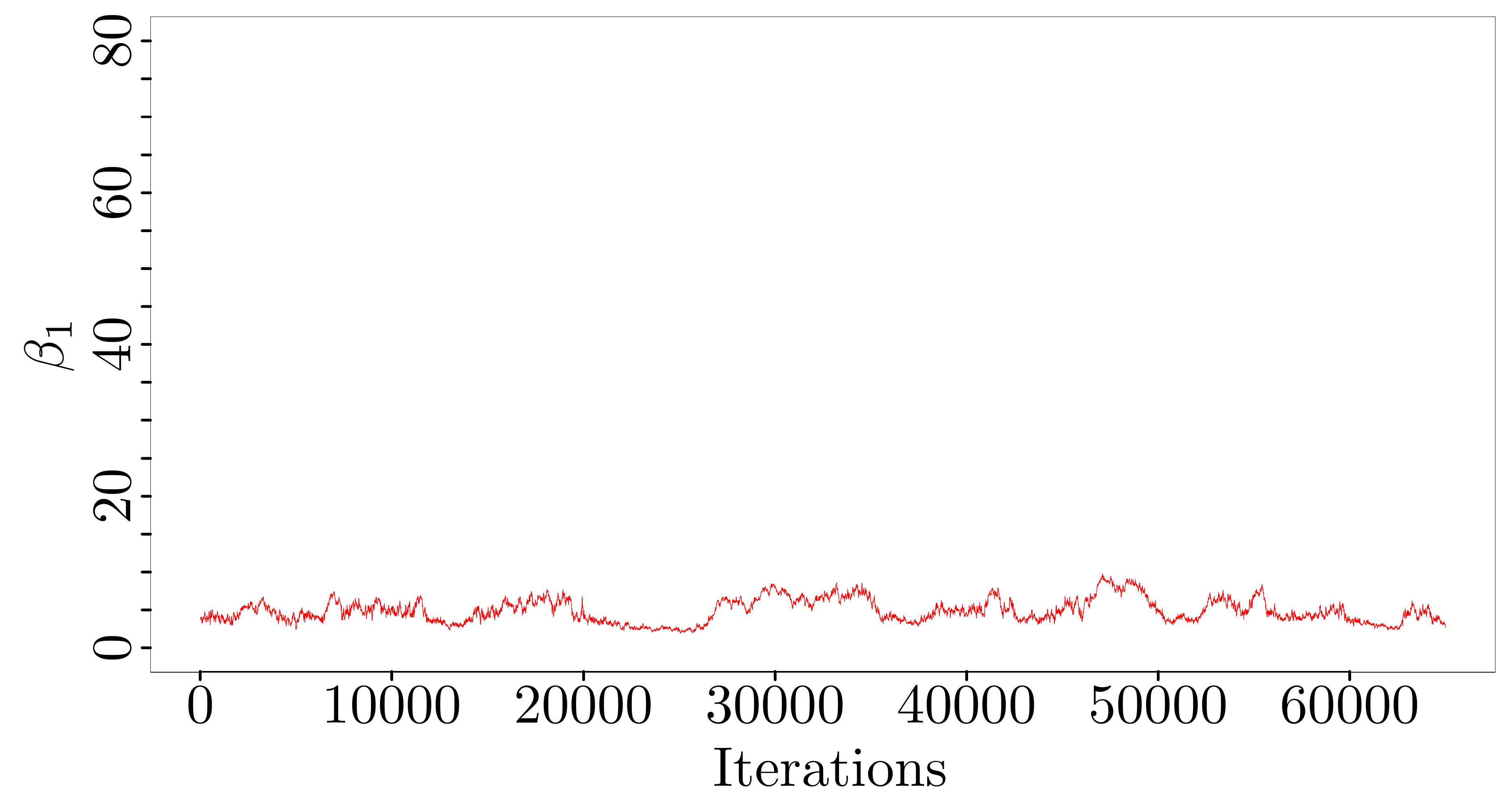}
\caption{Iteration history for parameter $\beta_1$. Each frame corresponds to an independent chain.}
\label{lack_of_conv}
\end{centering}
\end{figure}

The proposed method is used with $N=10^5$, $\alpha=.99$, $c=.1$, and $h_{max}=.1$. Five independent chains are generated from overdispersed initial iterates using adaptive MCMC \cite{vihola}. When $\nu=\lambda=0$, the MCMC sampler fails to converge because the tail of $p$ is too heavy. To illustrate this, Figure~\ref{lack_of_conv} shows the iteration history for the parameter $\beta_1$ in each of the five chains after a burn-in period is discarded. The two leftmost frames indicate that $p$ is heavy tailed; the other
three chains never reach the tail. A heavy-tailed PDF such as this
requires extensive sampling, which makes the reliable computation of
sensitivity indices intractable. Therefore, we use regularization to
alleviate this problem by increasing $\nu$ as we monitor the
sampler's convergence. We find that $\nu=.2$
yields converged chains with $N=10^5$ samples. The chains are deemed
convergent by using the potential scale reduction factor (PSRF) \cite{mpsrf} as well as visualizing the iteration histories and histograms from each of the five chains. 

\begin{figure}[h]
\begin{centering}
\includegraphics[width=.6 \textwidth]{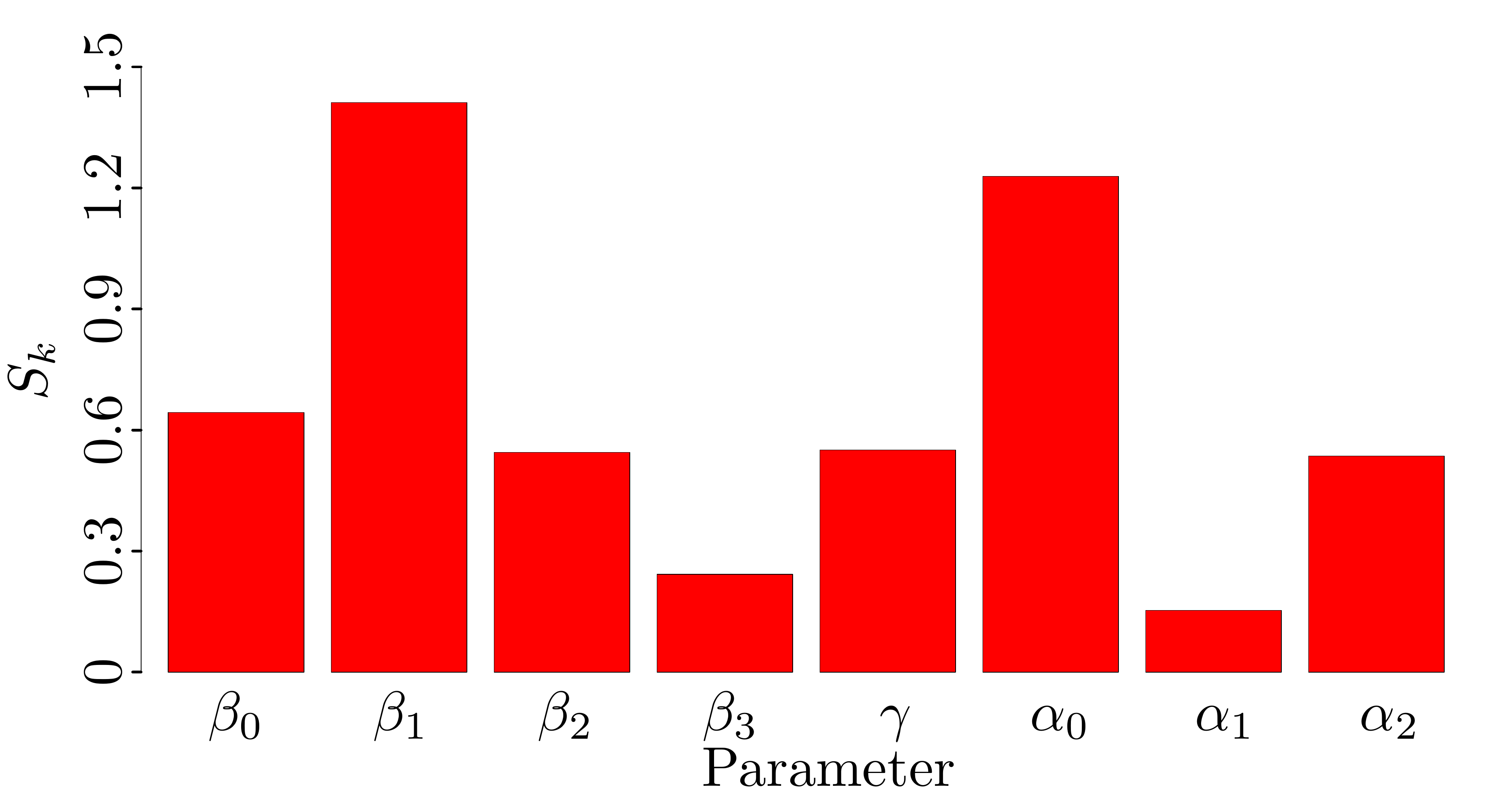}
\includegraphics[width=.39 \textwidth]{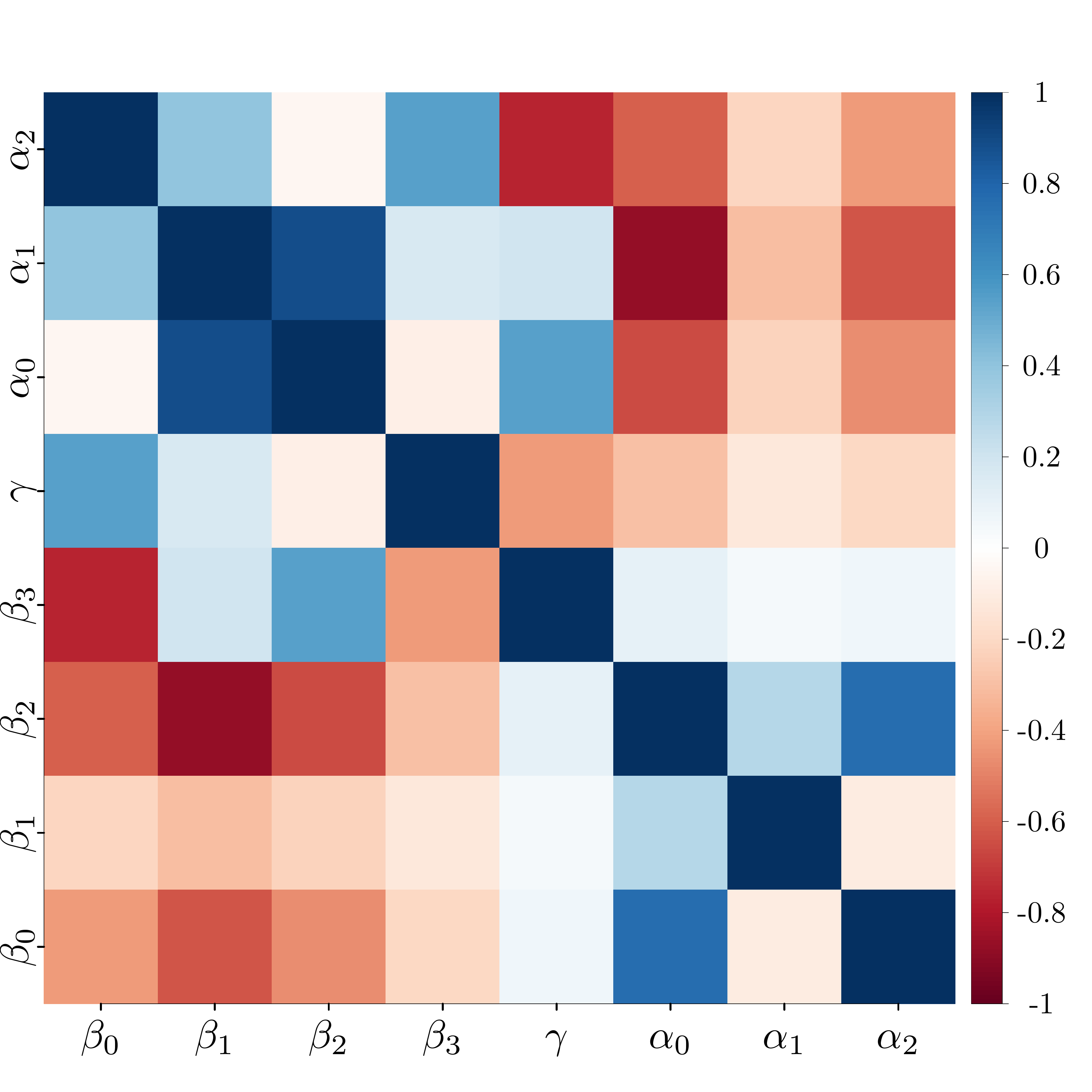}
\caption{Sensitivity indices (left) and Pearson correlation coefficients of the parameters (right) for the synthetic test problem.}
\label{syn_prob_results}
\end{centering}
\end{figure}

Plotting the iteration history of $\LL_\lambda$ indicates that a burn-in of $3.5 \times 10^4$ is sufficient. Then the remaining samples from the five chains are pooled together so that sensitivities and correlations may be computed from them.  Figure~\ref{syn_prob_results} shows the sensitivity indices and Pearson correlation matrix computed from the pooled samples. These results are consistent with our expectations, $\beta_1$ is seen as the most important parameter and $\alpha_1$ as the least important. Two primarily blocks are seen in the correlation plot representing the set of spatial variables and the set of temporal variables. Negative correlations are observed on the off diagonal blocks since the spatial and temporal variables are multiplied by one another and hence are inversely related.

\begin{figure}[h]
\begin{centering}
\includegraphics[width=.6 \textwidth]{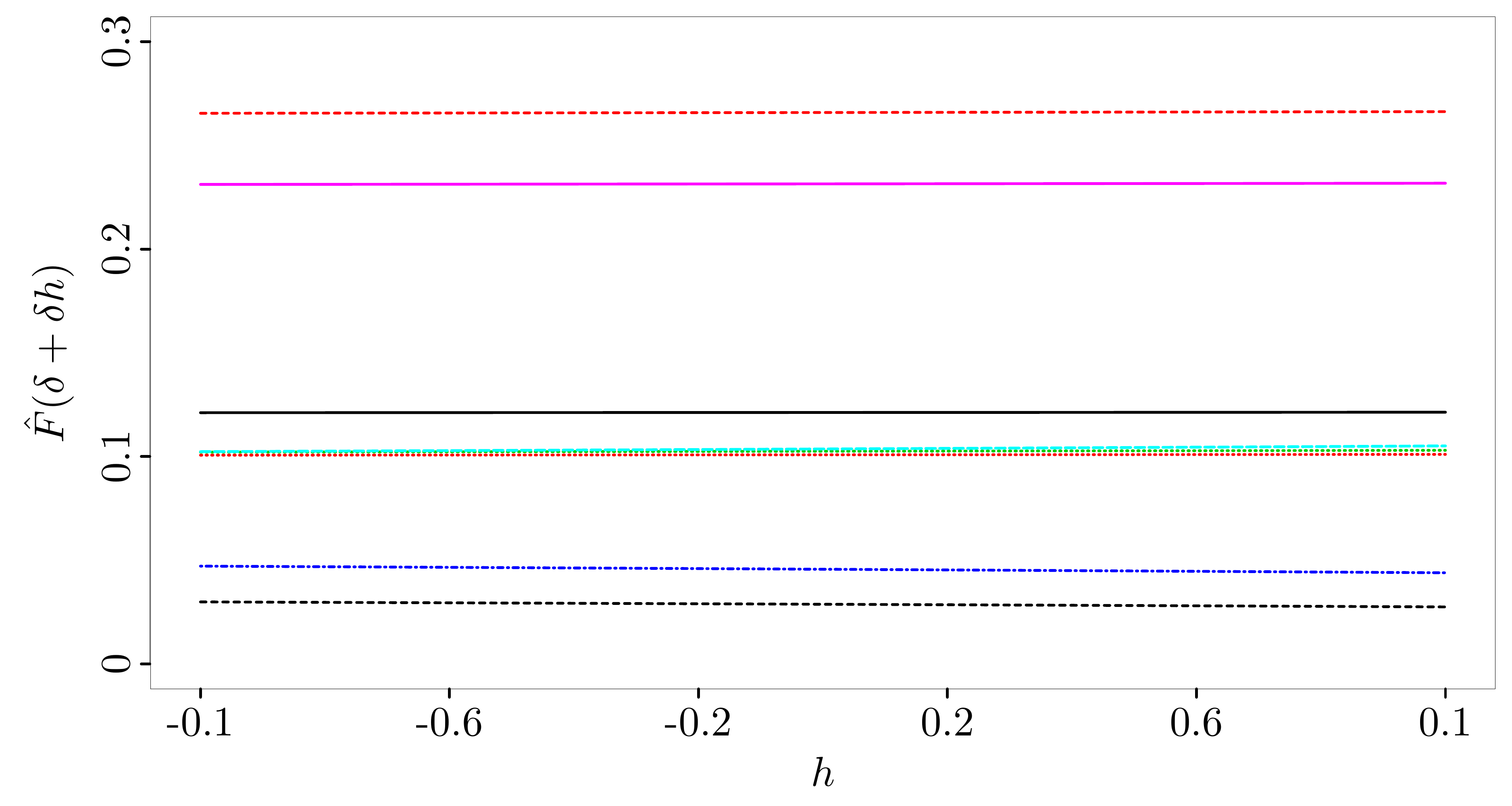}
\caption{Sensitivity index perturbations for the synthetic test problem. Each line corresponds to a given parameter.}
\label{syn_prob_pertubation}
\end{centering}
\end{figure}

Figure~\ref{syn_prob_pertubation} displays \eqref{Fhat} plotted for $h \in (-\frac{1}{10},\frac{1}{10})$, $k=1,2,\dots,8$. The horizontal lines indicate that errors in determining $\delta$ are immaterial since the analysis would be unchanged by perturbing $\delta$.

\subsection{Analysis of a space-time Gaussian process}
In this section we apply the proposed method to analyze the motivating statistical model \cite{bessac16}. The model aims at forecasting wind speed by fusing two heterogeneous datasets: numerical weather prediction model (NWP) outputs and physical observations. 
Let $\Ynwp$ denote the output of the NWP model and $\YObs$ denote the observed measurements. They are modeled as a bivariate space-time Gaussian process specified in terms of mean and covariance structures as follows,
\begin{align}
 \label{eq:GP:full}
 \begin{pmatrix}\YObs \\ \Ynwp  \end{pmatrix} \sim \mathcal{N}\left( \begin{pmatrix} \mu_{\rm{Obs}}(\theta) \\ \mu_{\rm{NWP}}(\theta) \end{pmatrix}  , \begin{pmatrix} \Sigma_{\rm{Obs}}(\theta)  & \Sigma_{\rm{Obs},\rm{NWP}}(\theta) \\   \Sigma_{\rm{Obs},\rm{NWP}}^{T}(\theta)& \Sigma_{\rm{NWP}}(\theta)
\end{pmatrix}\right), 
\end{align}
where $\theta$ is the set of parameters that describe the shapes of the means and covariances.  
The model is expressed in a hierarchical conditional manner to avoid the specification of the full joint covariance in \eqref{eq:GP:full}, indeed the mean and covariance of the distributions $\left( \YObs | \Ynwp  \right) \sim \mathcal{N}\Big( \mu_{Obs|NWP}, \Sigma_{Obs|NWP} \Big)$  and $\Ynwp  \sim \mathcal{N}\Big( \mu_{NWP}, \Sigma_{NWP} \Big)$ are specified in time, geographical coordinates, and parameters from the numerical model (the land-use parameter). 
More precisely, 
\begin{align}
  \label{eq:EqmuNWP}
\mu_{NWP}(t,s)  &= \left(\alpha_{0}\rm{LU}(s) + \alpha_{1}{\rm Lat}(s) + \alpha_{2}{\rm Long}(s)\right) f(t) ,
\end{align}
where $s$ is a spatial location, $t$ is time, $f(t)$ represents a sum of temporal harmonics with daily, half-daily and 8hr-periodicities, $\rm{LU}(s)$ is a categorical variable that represents the land-use associated with location $s$, ${\rm Lat}$ and ${\rm Long}$ are the latitude and longitude coordinates. 
\begin{align}\label{eq:EqCov}
\Sigma_{NWP}(.,s_{i};.,s_{j}) = \C(\Ynwp(.,s_{i}),\Ynwp(.,s_{j})) = (\Psi(s_{i}) \Gamma_{0}  \Psi(s_{j})^{T}) + \delta_{i-j} \Gamma_{LU(s_{i})},
\end{align}
$\Gamma_{0}$, $(\Gamma_{LU(s_{i})})_{i=1..I}$ are temporal squared exponential covariances expressed as $$\Gamma_{.}(t_k,t_l) = \sigma_{.} \exp(- \rho_{.} (|t_k - t_l|)^{2}) +  \delta_{k-l} \gamma_{.},$$
where $\delta_{k-l}$ is the Kronecker delta, $\sigma$, $\rho$, and $\gamma$ are parameters to be estimated. 
$(\Gamma_{LU(s_{i})})_{i=1..I}$ are land-use specific terms, and $\Psi$ is linear in the latitude and longitude coordinates and quadratic in time. 
The parameters $\alpha_{0}$, $\alpha_{1}$, $\alpha_{2}$, along with the parameters of $f(t)$, $(\Gamma_{s_i})_{i=0..I}$ and $\Psi$, will be estimated during the maximum likelihood procedure. We will denote the collection of all these parameters by $\theta_{NWP}$.

The conditional distribution is expressed through its mean and covariance: 
\begin{align}
\nonumber \mu_{Obs|NWP}(t,s) &= \mu(t,s) + (\Lambda \Ynwp)(t,s)\,, \\  
\end{align}
where $\mu(t,s)$ is written similarly to $\mu_{NWP}(t,s)$ as a product of temporal harmonics and a linear combination of the coordinates latitude and longitude. $\Lambda$ is a projection matrix specified in time, latitude, longitude and the land-use parameter. 
The covariance $\Sigma_{Obs|NWP}$ is parametrized with a similar shape to \eqref{eq:EqCov} with a different set of parameters. 
Parameters of these functions are denoted as $\theta_{Obs\vert NWP}$ in the following and will estimated by maximum likelihood.

This model is fitted by maximum likelihood on the two datasets with respect to the parameters $\theta=(
\theta_{NWP},\theta_{Obs \vert NWP})$.  
The negative log likelihood of the model can be decomposed as
\begin{eqnarray}
\LL(\theta)=\LLnwp(\theta_{NWP})+\LLobs(\theta_{Obs \vert NWP}),
\end{eqnarray}
where $\LLnwp(\theta_{NWP})$ and $\LLobs(\theta_{Obs\vert NWP})$ are the negative log likelihoods for the marginal distribution of $\Ynwp$ and the conditional distribution $\YObs \vert \Ynwp$, respectively. Since the model decomposes in this way, we will consider analysis of the parameters in $\Ynwp$ and $\YObs \vert \Ynwp$ separately. Our dataset consists of 27 days of measurements from August 2012; details may be found in \cite{bessac16}. The parameter sensitivity during the first 13 days for $\Ynwp$ and $\YObs \vert \Ynwp$ is analyzed in Subsection~\ref{nwp_numerics} and Subsection~\ref{cond_numerics}, respectively. Inferences are drawn from this analysis and validated using the later 14 days of data in Subsection~\ref{validation}.

\subsubsection{Parameter sensitivity analysis for $\Ynwp$}
\label{nwp_numerics}
In this subsection we apply the proposed method to determine the sensitivity of the marginal model for $\Ynwp$ to its 41 parameters during a 13-day period. The sensitivities being computed are with respect to the parameters in $\theta_{NWP}$, but for notational simplicity we will denote them by $\theta$ in this subsection.

In order to determine $\theta^{\star}$ (line~1 of Algorithm~\ref{alg:aux}), the
L-BFGS-B algorithm is used to minimize $\LLnwp$. Visualizing the model
predictions for various choices of $c$ yields $c=.35$ (line 2 of
Algorithm~\ref{alg:aux}). Then $M$ \eqref{M} is estimated with
$\ell=5000$ Monte Carlo samples (line 3 of
Algorithm~\ref{alg:aux}). It returns an estimate $M=4160$ with
standard deviation 2; hence $\ell$ is considered to be sufficiently
large. These steps complete the preprocessing stage by providing characteristic values for the parameters $\theta$ and the loss function $\LL$.

The PDF $p$ is found to be heavy tailed, so $\nu=.1$ is chosen
to reduce this effect.
Then the equation $\Delta(\delta)=\alpha$ is solved with $\alpha=.99$ by evaluating $\Delta$ and manually updating $\delta$. The solution $\delta=.07$ is obtained. This converged in very few iterations because of the nice properties of the equation $\Delta(\delta)=\alpha$. Any other derivative-free nonlinear solver may be used in our framework; however, manual tuning is preferable in many cases because of the simplicity of the equation and the stochasticity of the function evaluations. Having determined $\nu$ and $\delta$, the PDF $p$ from which we draw samples is now well defined.

Adaptive MCMC \cite{vihola} is used with a desired acceptance rate of
$.15$. Five chains of length $N=4 \times 10^5$ each are generated
independently from overdispersed initial iterates, and the first
$10^5$ iterates are discarded as burn-in. The PSRF convergence diagnostic from \cite{mpsrf} is used on $\theta_k$ and $\frac{\partial \LL}{\partial \theta_k}$ separately to assess the convergence of each. The PSRFs for all parameters lie in the intervals $(1,1.025)$ and $(1,1.048)$, respectively. Other visual diagnostics are applied as well, along with comparing sensitivity indices from each of the chains. The sensitivity estimation appears to converge.

\begin{figure}[h]
\begin{centering}
\includegraphics[width=.95 \textwidth]{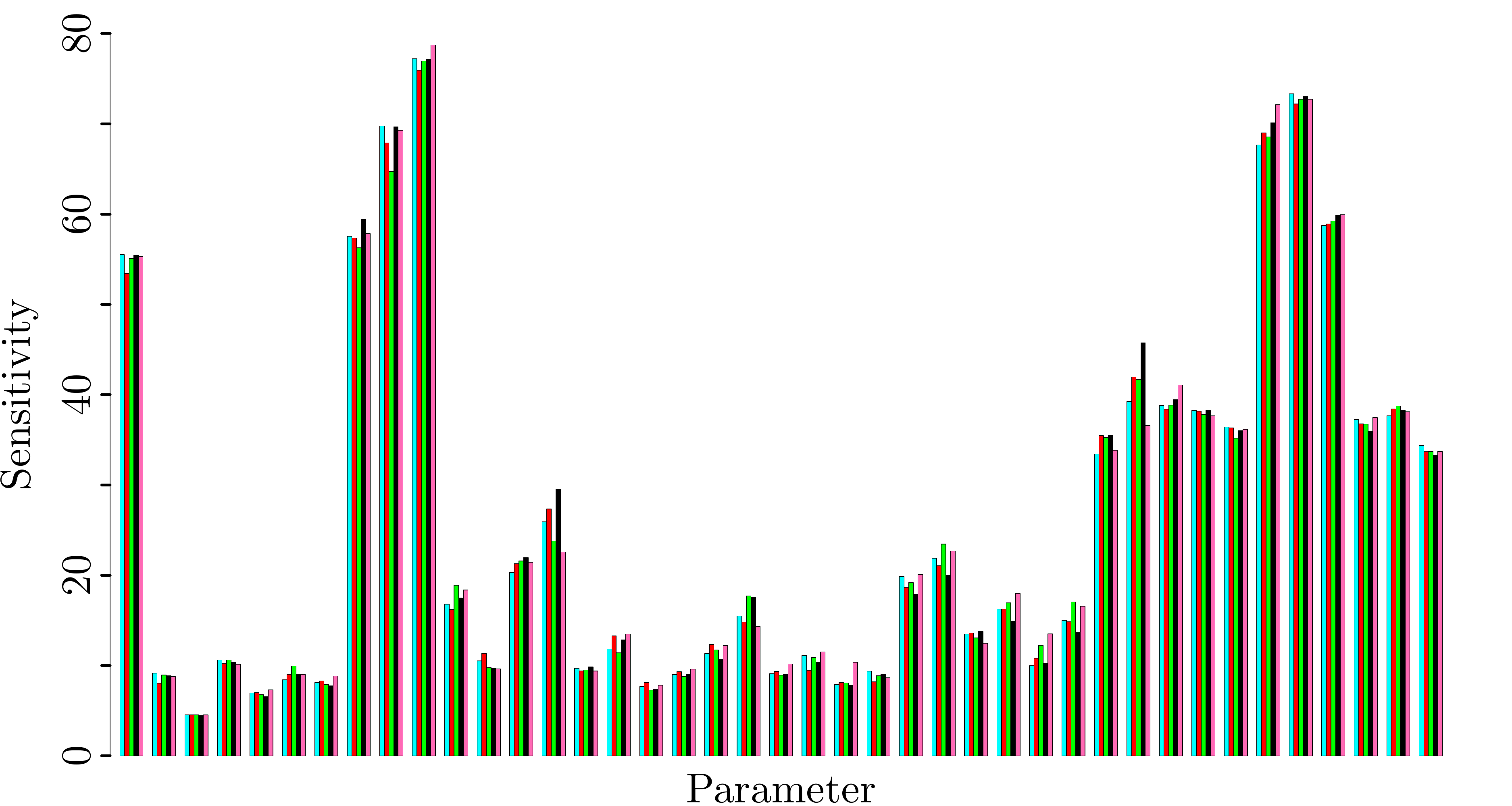}
\caption{Sensitivity indices for $\Ynwp$. The five colors represent
  the sensitivity indices computed from each of the five
  chains.}
\label{nwp_sensitivities}
\end{centering}
\end{figure}

Figure~\ref{nwp_sensitivities} displays the sensitivity indices estimated from each of the chains. The five different colors represent the five different chains; their comparability demonstrates that MCMC estimation errors are negligible. The intercept terms in the mean and the covariance kernel parameters are the most influential. The terms parameterizing $\Psi$ are less influential, particularly the quadratic temporal terms.

\begin{figure}[h]
\begin{centering}
\includegraphics[width=.65 \textwidth]{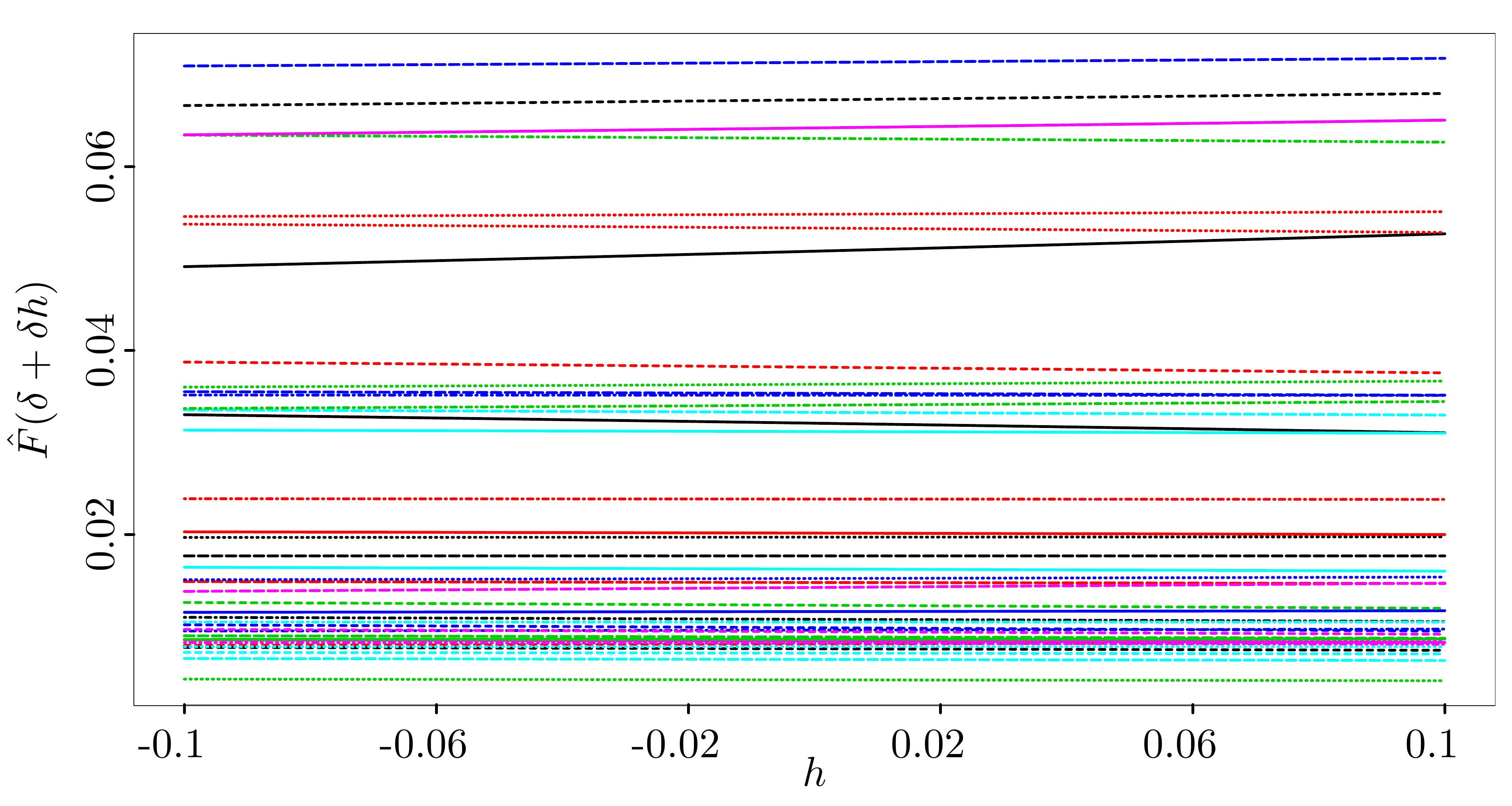}
\caption{Sensitivity index perturbations for $\Ynwp$. Each line corresponds to a given parameter.}
\label{nwp_pertubation}
\end{centering}
\end{figure}

As discussed in Subsection~\ref{delta_sensitivity}, the robustness of the sensitivities with respect to errors in $\delta$ may be estimated as a by-product of computing sensitivities. Figure~\ref{nwp_pertubation} displays \eqref{Fhat} plotted for $h \in (-1/10,1/10)$, $k=1,2,\dots,41$. Most of the curves are nearly horizontal, and those that not horizontal display small variation that does not change the resulting inference. Thus the sensitivities are robust with respect to $\delta$, and hence any errors made when determining $\delta$ are negligible. 

\begin{figure}[h]
\begin{centering}
\includegraphics[width=.65 \textwidth]{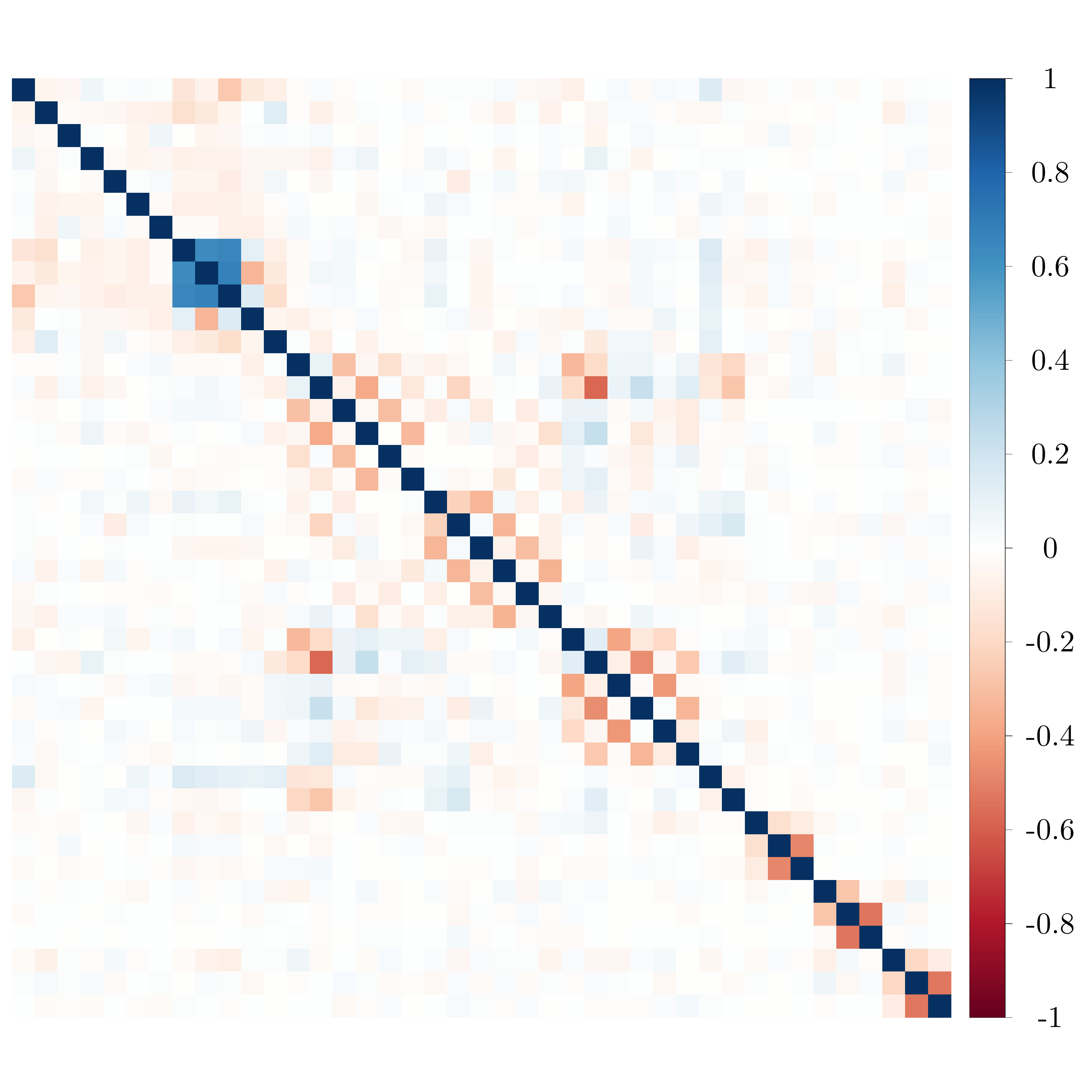}
\caption{Pearson correlation coefficients for the parameters of
  $\Ynwp$. }
\label{nwp_cor}
\end{centering}
\end{figure}

As mentioned in Subsection~\ref{correlations}, parameter correlation information is a useful complement to sensitivity indices. Figure~\ref{nwp_cor} displays the empirical Pearson correlation matrix computed from the $1.5 \times 10^6$ MCMC samples retained after removing burn-in and pooling the chains. Strong positive correlations are observed between the three land-use dependent spatial intercepts in the mean. Strong negative correlations are observed between the temporal range $\rho$ and the nugget term $\gamma$ parameterizing the land-use specific covariance kernels $\Gamma_{LU(s_i)}$. This correlation is expected since the nugget term represents the variance of the signal that is not explained by the exponential part. 

\subsubsection{Parameter sensitivity analysis for $\YObs \vert \Ynwp$}
\label{cond_numerics}
In this subsection we apply the proposed method to determine the sensitivity of the model for $\YObs \vert \Ynwp$ to its 54 parameters during the same 13-day period used in Subsection~\ref{nwp_numerics}. The sensitivities being computed are with respect to the parameters in $\theta_{Obs \vert NWP}$, but for notational simplicity we will denote them by $\theta$ in this subsection.

In a similar fashion to Subsection~\ref{nwp_numerics}, the L-BFGS-B algorithm is used to determine $\theta^\star$ (line~1 of Algorithm~\ref{alg:aux}). Visualizing the model predictions for various choices of $c$ yields $c=.05$ (line 2 of Algorithm~\ref{alg:aux}). Then M \eqref{M} is estimated with $\ell=5000$ Monte Carlo samples (line 3 of Algorithm~\ref{alg:aux}). It returns an estimate $M=2973$ with standard deviation 2; hence $\ell$ is considered to be sufficiently large. These steps complete the preprocessing stage by providing characteristic values for the parameters $\theta$ and the loss function $\LL$. One may note that $c$ and $M$ are significantly smaller for $\YObs \vert \Ynwp$ than for $\Ynwp$. Their difference is unsurprising since $\Ynwp$ and $\YObs \vert \Ynwp$ model two different processes.

The PDF $p$ is found to be heavy tailed, so $\nu=.15$ is chosen to reduce the tail of $p$. Analogously to Subsection~\ref{nwp_numerics}, $\Delta(\delta)=\alpha=.99$ is solved yielding $\delta=.06$ (line~3 of Algorithm~\ref{alg:sampling}). 

Adaptive MCMC \cite{vihola} is used with a desired acceptance rate of $.15$. Five chains of length $N=4 \times 10^5$ each are generated independently from overdispersed initial iterates, and the first $10^5$ iterates are discarded as burn-in. The convergence diagnostic from \cite{mpsrf} is used on $\theta_k$ and $\frac{\partial \LL}{\partial \theta_k}$ separately to assess the convergence of each. The PSRFs for all parameters lie in the intervals $(1,1.191)$ and $(1,1.036)$, respectively. Other visual diagnostics are applied as well, along with comparing sensitivity indices from each of the chains. A few sensitivity indices have not fully converged; however, the remaining uncertainty in their estimation is sufficiently small for our purposes. These uncertain sensitivities are among the largest in magnitude. Since our goal is encouraging model parsimony, then precisely ordering the most influential parameters is of secondary importance.

\begin{figure}[h]
\begin{centering}
\includegraphics[width=.95 \textwidth]{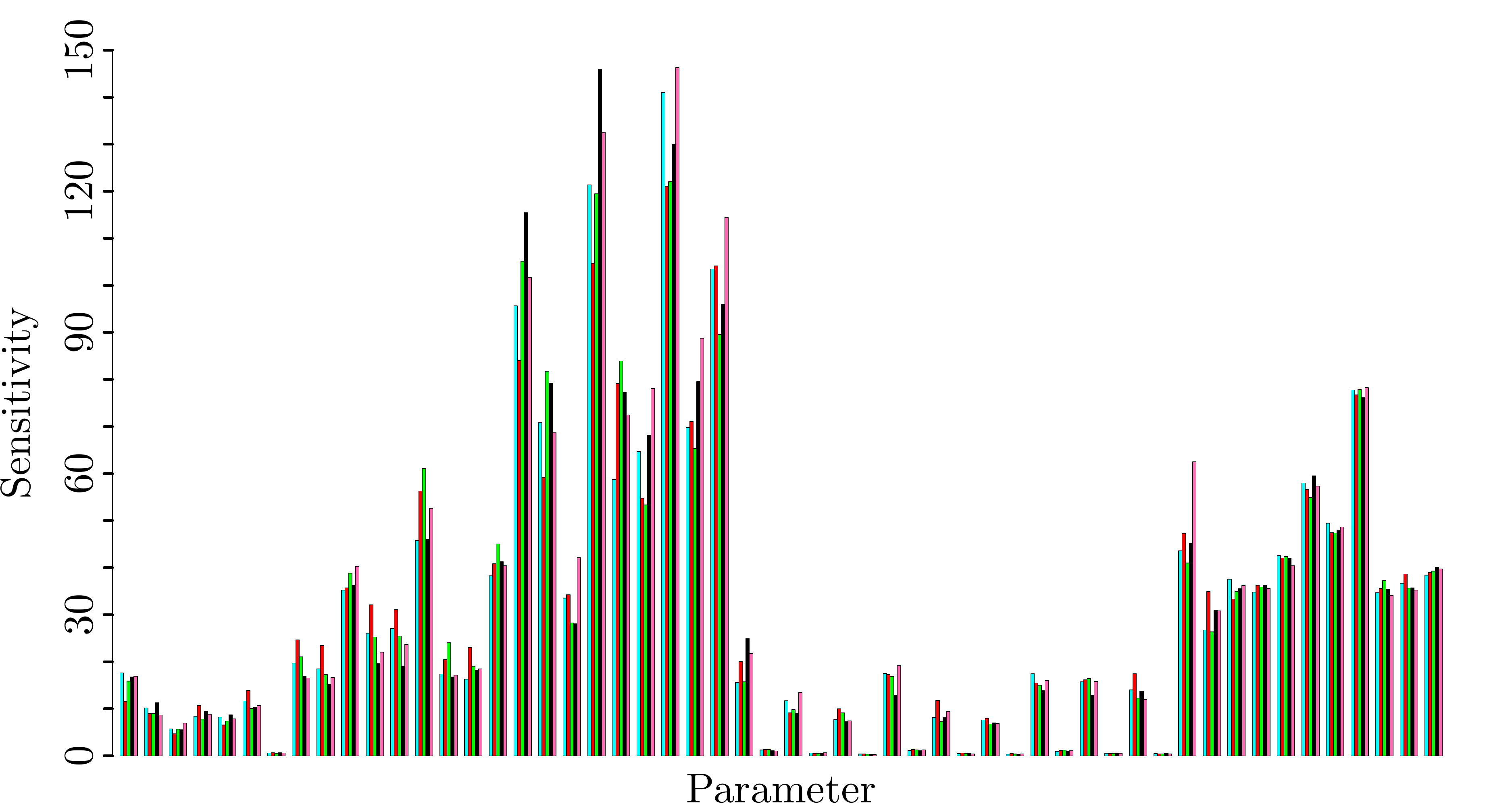}
\caption{Sensitivity indices for $\YObs \vert \Ynwp$. The five colors represent the sensitivity indices computed from each of the five chains.}
\label{cond_sensitivities}
\end{centering}
\end{figure}

Figure~\ref{cond_sensitivities} displays the sensitivity indices estimated from each of the chains. The five different colors represent the five different chains. The sensitivities with greatest uncertainties are demonstrated by the differences in their estimated values in each chain; however, these discrepancies are sufficiently small that they do not alter our resulting inference. The longitudinal terms in the mean and $\Psi$ are observed to have little influence since their sensitivity indices are nearly zero. The most influential parameters are the spatial weights in the matrix $\Lambda$ which acts on $\Ynwp$.

\begin{figure}[h]
\begin{centering}
\includegraphics[width=.65 \textwidth]{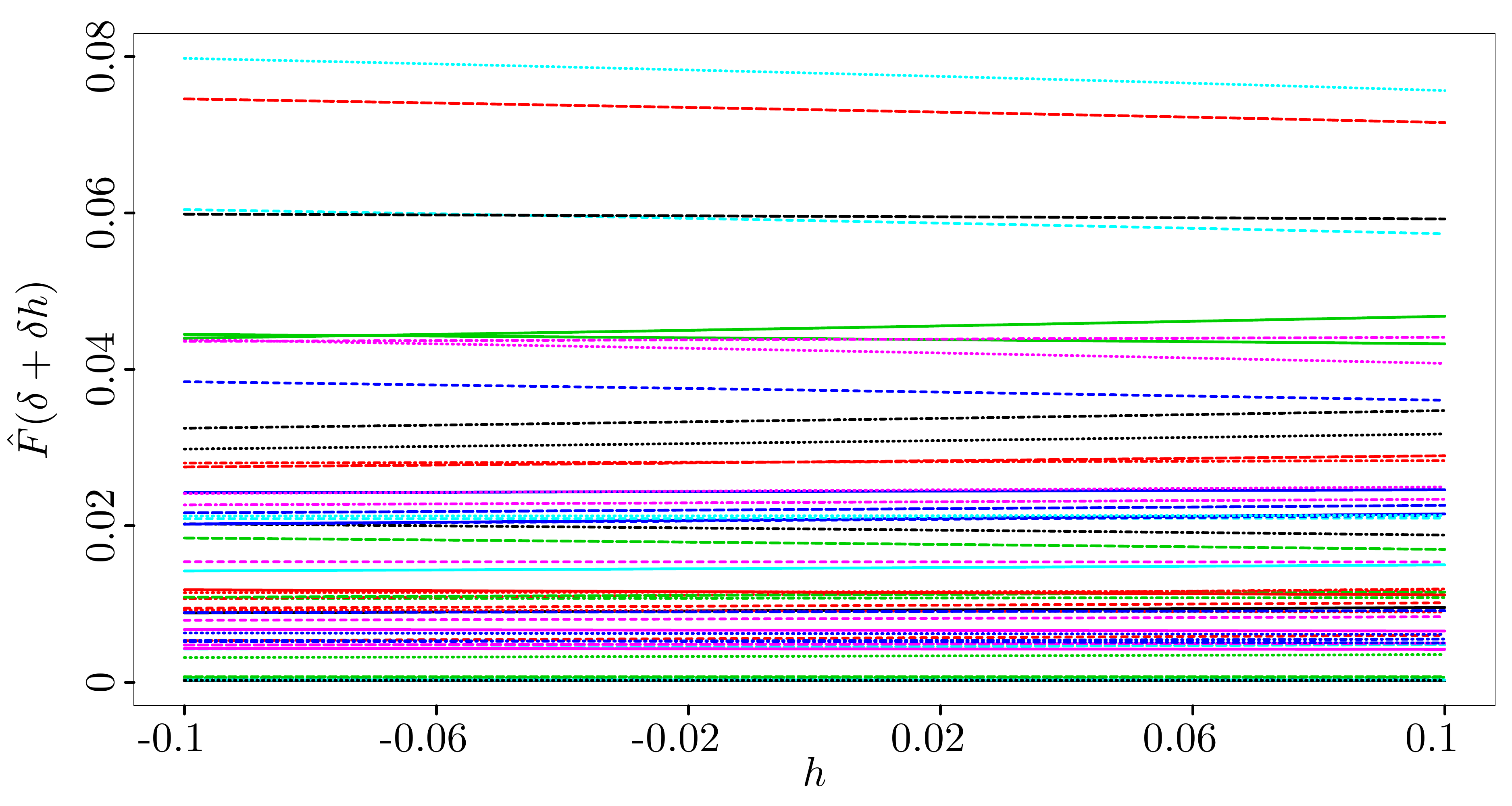}
\caption{Sensitivity index perturbations for $\YObs \vert \Ynwp$. Each line corresponds to a given parameter.}
\label{cond_pertubation}
\end{centering}
\end{figure}

As discussed in Subsection~\ref{delta_sensitivity}, the robustness of the sensitivities with respect to errors in $\delta$ may be estimated as a by-product of computing sensitivities. Figure~\ref{cond_pertubation} displays \eqref{Fhat} plotted for $h \in (-\delta/10, \delta /10)$, $k=1,2,\dots,54$. Most of the curves are nearly horizontal, and those that are not horizontal display small variation that does not change the resulting inference.

\begin{figure}[h]
\begin{centering}
\includegraphics[width=.65 \textwidth]{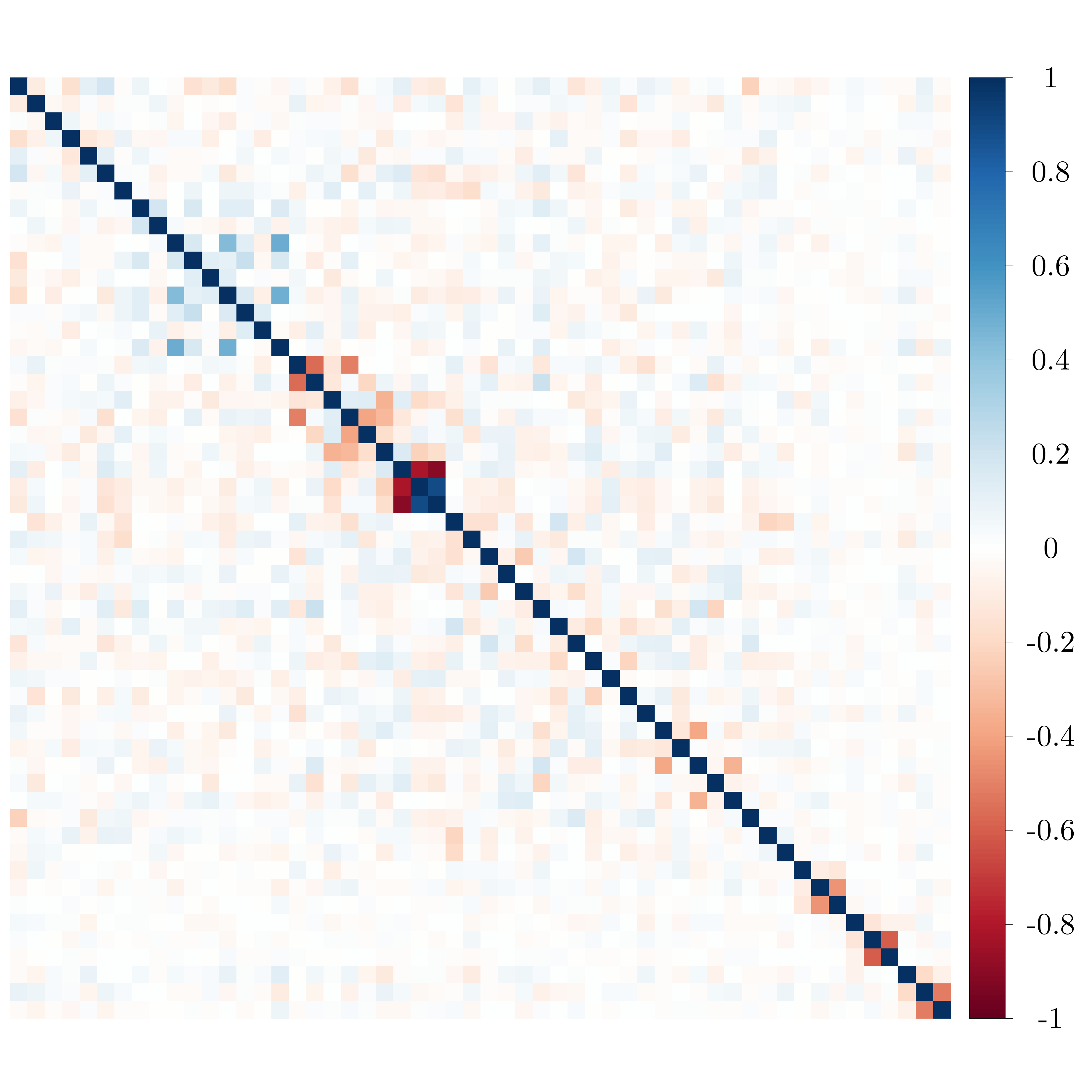}
\caption{Pearson correlation coefficients for the parameters of $\YObs \vert \Ynwp$.}
\label{cond_cor}
\end{centering}
\end{figure}

As mentioned in Subsection~\ref{correlations}, parameter correlation information is a useful complement to sensitivity indices. Figure~\ref{cond_cor} displays the empirical Pearson correlation matrix computed from the $1.5 \times 10^6$ MCMC samples retained after burn-in and pooling the chains. Strong correlations are observed between the spatial weights in the matrix $\Lambda$. Similar to $\Ynwp$, strong negative correlations are also observed between the temporal range $\rho$ and the nugget term $\gamma$ of the land-use specific covariance kernels. 

\subsubsection{Inference and validation of results}
\label{validation}
In some cases with mathematical models one may set a threshold and fix all parameters whose sensitivity is below the
threshold. This approach is not suitable for statistical models
because the parameters must be understood in light of their
contribution to the model structure and their correlation with other
parameters. Rather, the sensitivity indices and correlation structures should be used
to re-parameterize the statistical model in a simpler way. For instance, if a collection of spatially dependent
parameters are all found to be unimportant then the user may consider replacing
them by a single parameter which is not spatially dependent. 

Using the results of Subsection~\ref{nwp_numerics}, and
considerations of the model structure, we determine that the $\Ynwp$
model is insensitive to the temporal quadratic terms in the
parameterization of $\Psi$. Similarly, coupling the
results of Subsection~\ref{cond_numerics}, and the model structure, we
determine that the $\YObs \vert \Ynwp$ model is insensitive to
several of the longitude terms. Specifically, the longitude term in
the parameterization of the mean and the nine longitude terms in the
parameterization of $\Psi$. This conclusion confirms what one
would expect from the physics considerations. The flow is
predominantly east-west and thus the north-south correlation is
relatively weaker. 
The east-west information is likely to be well captured by $\Ynwp$ and hence is not needed in $\YObs \vert \Ynwp$. 

The 16 insensitive parameters are removed from the model, yielding a more parsimonious model that we refer to as the reduced model. To validate our inferences, we use the original model and the reduced model for prediction on the other 14 days of data we have available but did not use in the sensitivity analysis. Leave-one-out cross-validation is used to fit each of the models and assess their predictive capabilities.

\begin{figure}[h]
\begin{centering}
\includegraphics[width=.49 \textwidth]{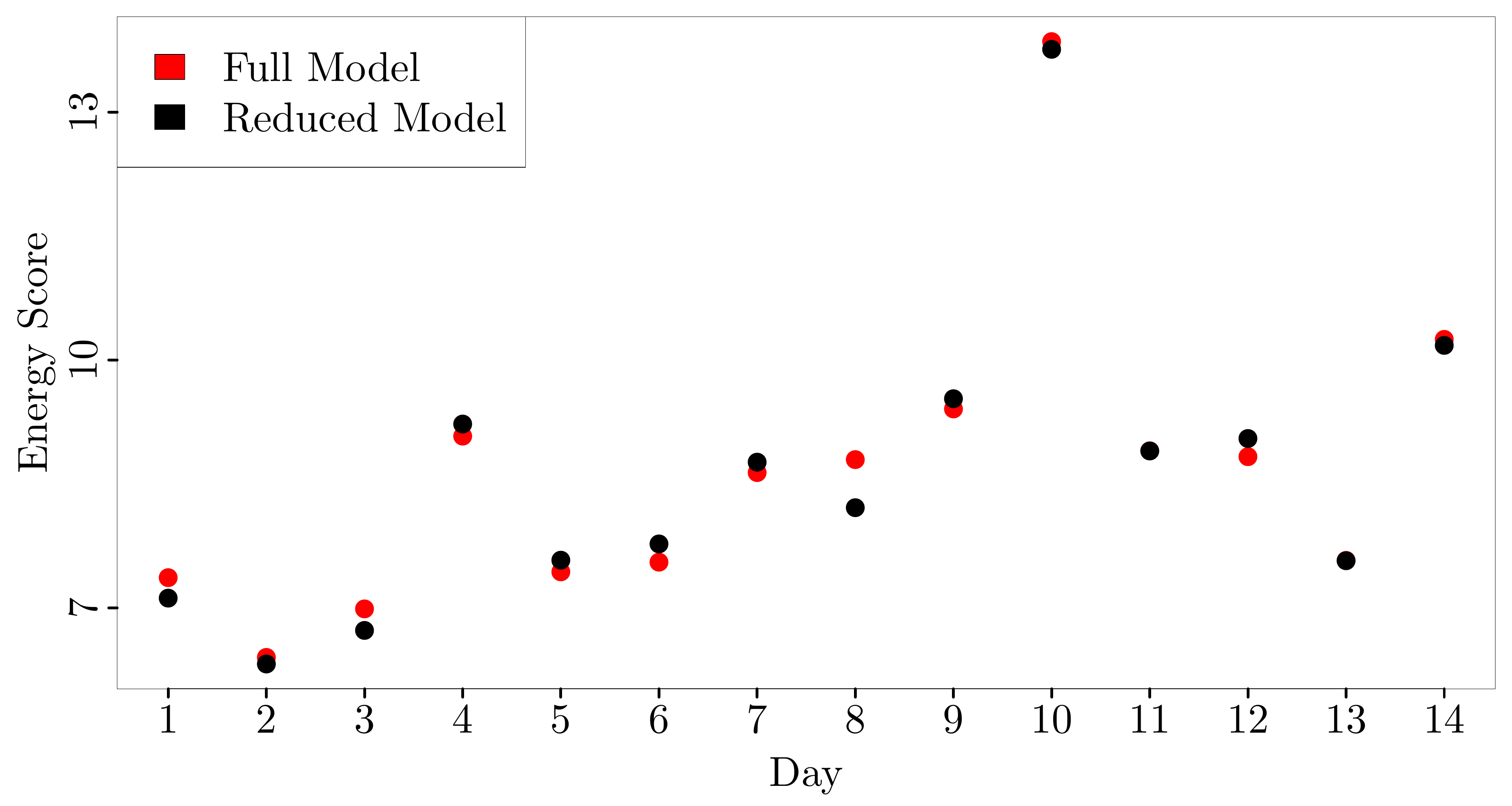}
\includegraphics[width=.49 \textwidth]{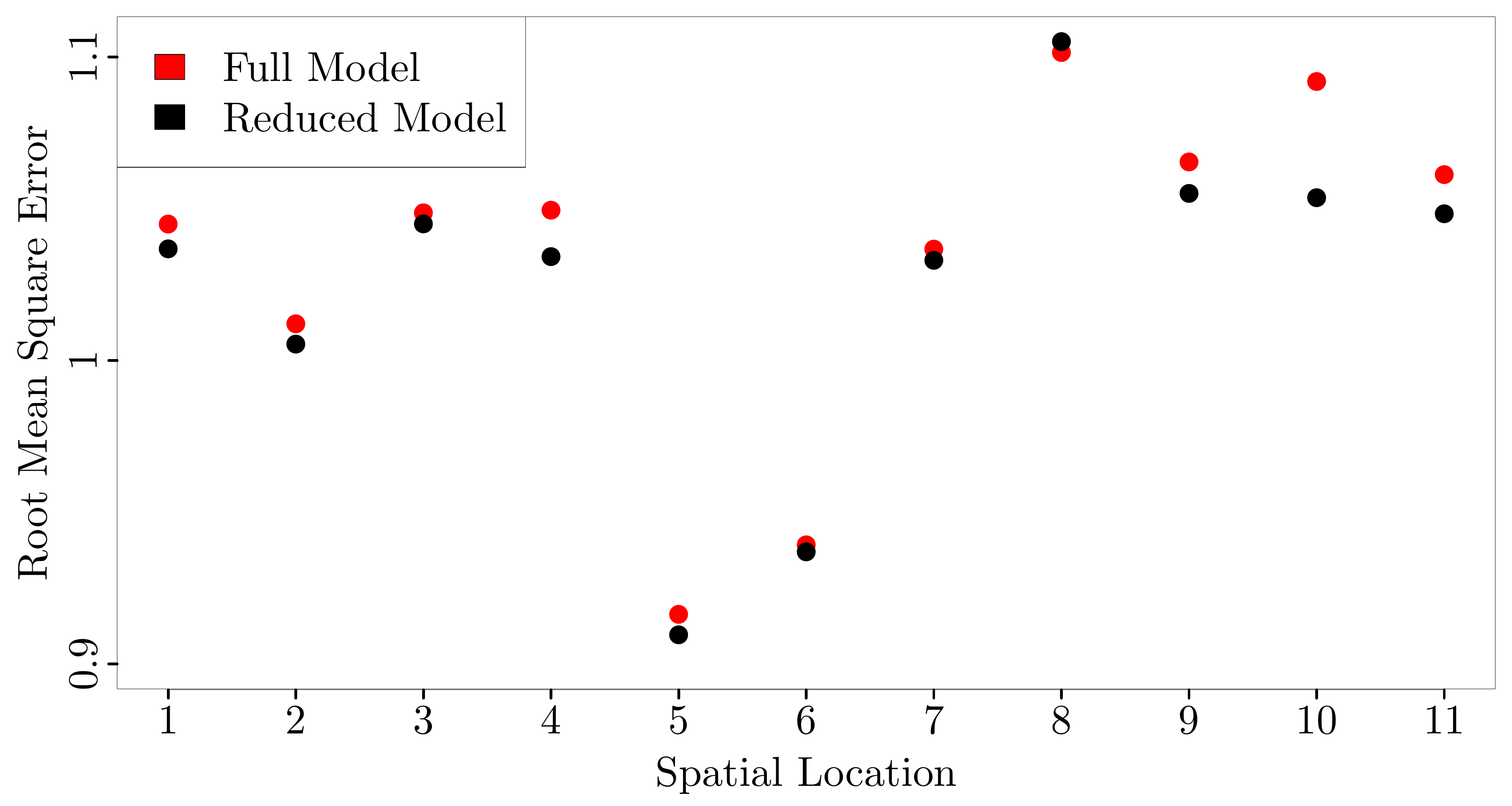}
\caption{Left: energy scores for each of the 14 days being predicted; right: root mean square error for each of the 11 spatial locations being predicted. The full model is red and the reduced model is black. }
\label{validation_comparison}
\end{centering}
\end{figure}

We simulated 1,000 scenarios $\Ysim$ for each of the 14 days and use two metrics to quantify the predictive capacity of the full and reduced models, namely, the energy score and the root mean square error. The energy score\footnote{$\displaystyle ES(\Ysim,\YObs)=\frac{1}{N}\sum_{i=1}^{N}|| Y_{\rm sim}^{(i)} - \YObs || - \frac{1}{2N^2}\sum_{i,j=1}^{N}|| Y_{\rm sim}^{(i)} - \tilde{Y}_{\rm sim}^{(j)} ||$, where $Y_{\rm sim}^{(i)}$ and $\tilde{Y}_{\rm sim}^{(i)}$ are independent predictive scenarios with $N=1,000$} \cite{es1,es2} is a measure of distance between the distribution used to generate the scenarios $\Ysim$ and the observed data $\YObs$ on a fixed day; hence 14 energy scores are computed (one for each day). The root mean square error\footnote{$\displaystyle RMSE(\Ysim,\YObs)=\sqrt{\frac{1}{T}\sum_{t=1}^{T}\left( \frac{1}{N}\sum_{i=1}^{N} \left( Y_{\rm sim}^{(i)}(t)-\YObs(t)\right)^2 \right)}$} is computed as the square root of the time average squared error between the average of scenarios and the observation at each spatial location; hence, there are 11 root mean square errors. Figure~\ref{validation_comparison} displays the energy scores on the left and root mean square errors on the right. The reduced model has slightly smaller (and hence better) energy scores and root mean square errors in the majority of the cases. The sum of energy scores for full model and reduced model is 121.1 and 120.7, respectively. The sum of root mean squared errors for the full model and reduced model is 11.4 and 11.3, respectively.

\begin{figure}[h]
\begin{centering}
\includegraphics[width=.49 \textwidth]{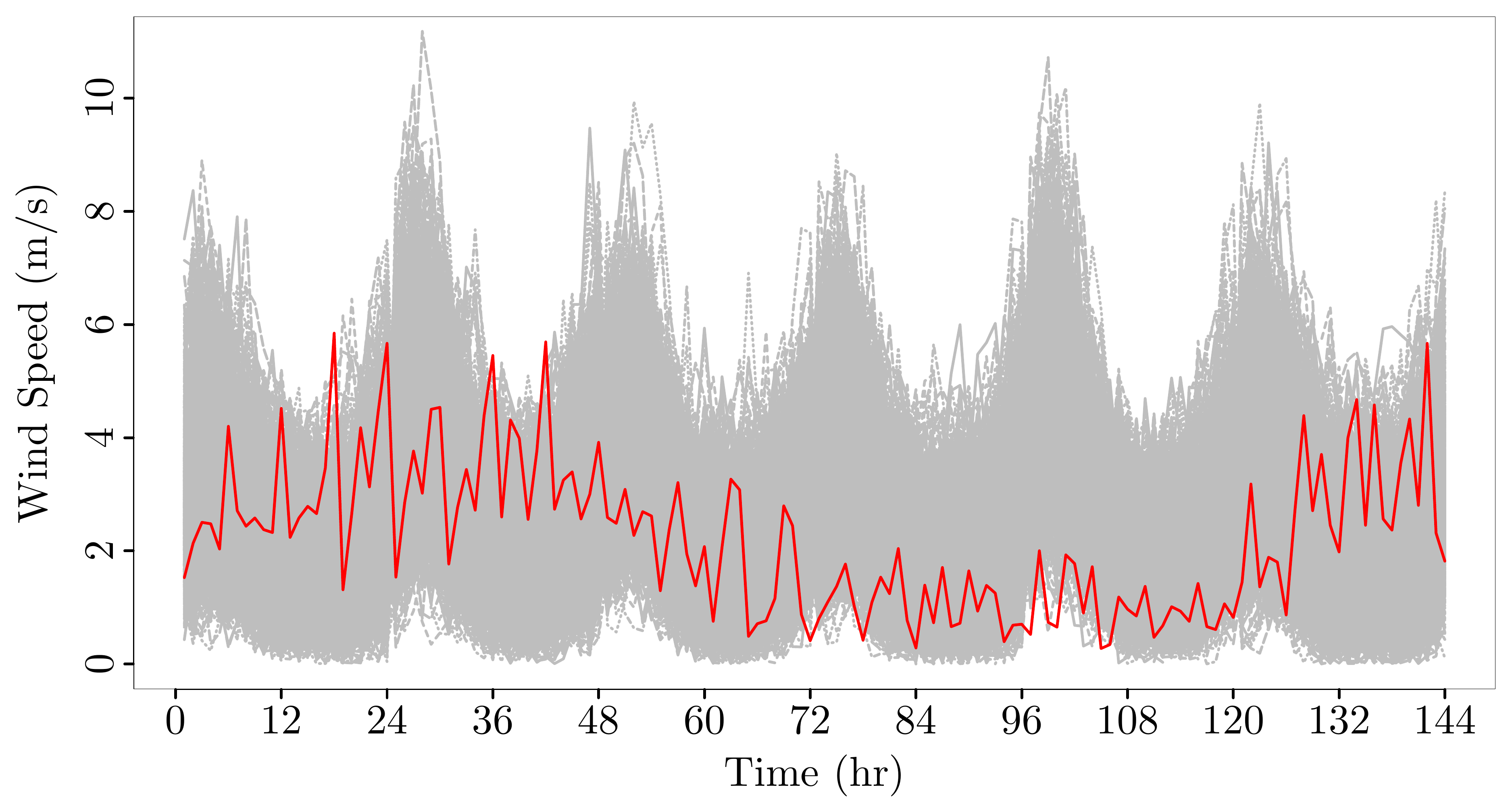}
\includegraphics[width=.49 \textwidth]{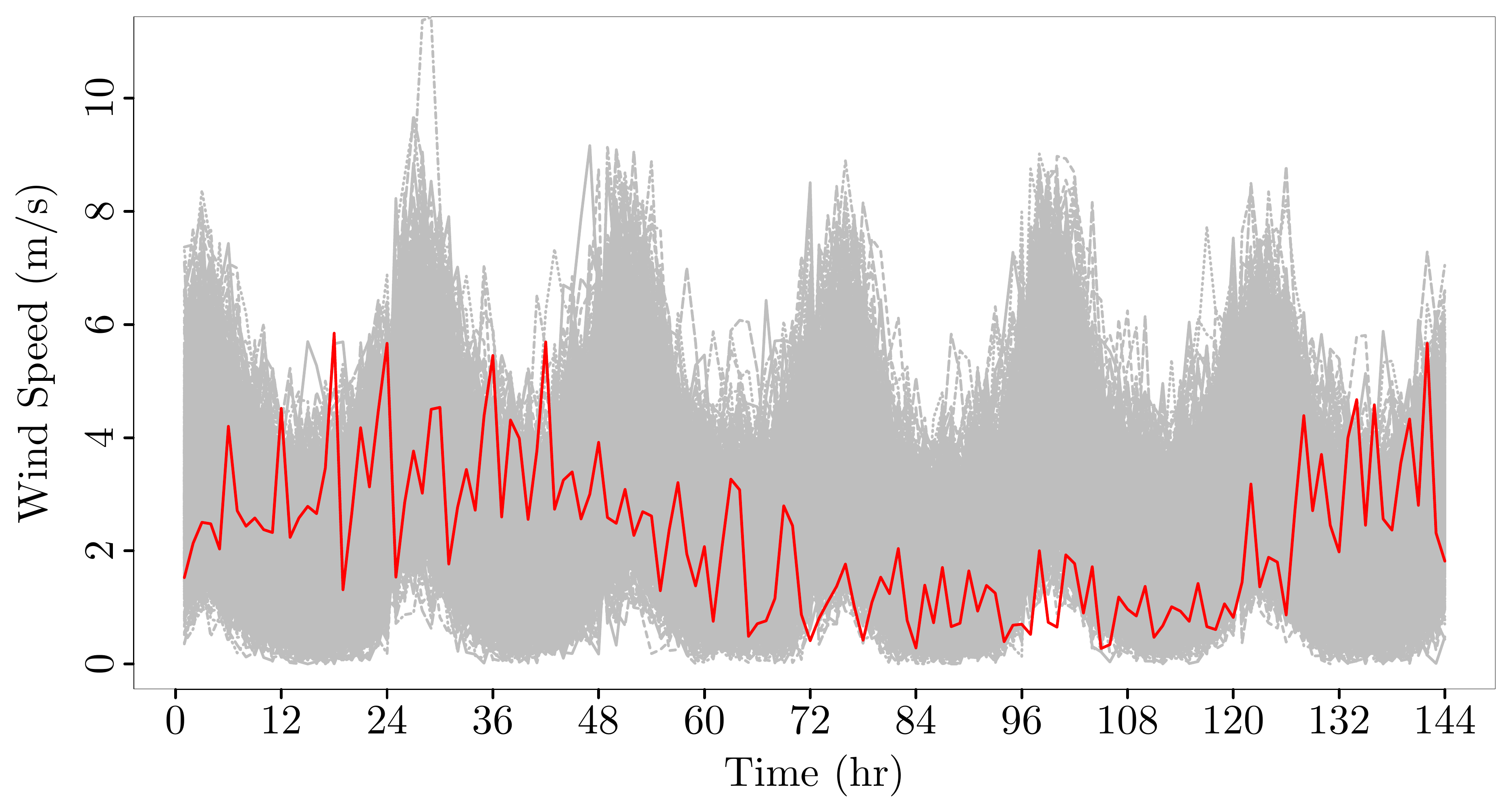}
\caption{Predictions using the full model (left) and reduced model
  (right) for 6 days at a fixed spatial location. The red
  curve is the observed wind speed and the grey curves are 1000
  simulations generated from each model.}
\label{validation_comparison_scenarios}
\end{centering}
\end{figure}

To further illustrate the difference between the full and reduced models, we display the simulated scenarios for a typical case. Specifically, we take the spatial location with median root mean square error and six days with median energy score and plot the 1,000 scenarios along with the observed data. Figure~\ref{validation_comparison_scenarios} displays the results with the full model on the left and reduced model on the right.

We have thus simplified the parameterization of the model from having 95 parameters to 79. The reduced model has equal or better predictive capability and is simpler to fit and analyze. Further, the reduced model typically has fewer outlying scenarios, as evidenced in Figure~\ref{validation_comparison_scenarios}. 

For this application we observed a strong insensitivity of $\YObs
\vert \Ynwp$ to some of its terms contributing longitudinal
information. This is likely because the $\Ynwp$ model captures
longitudinal information well and hence the Gaussian process does not need to fit terms contributing longitudinal information. Thus, inferences may also be made on the data being input to the statistical model through the parameter sensitivities. 

These inferences and simplifications are useful for multiple reasons. First, long term weather prediction is difficult so the parameters must be optimized frequently to accommodate changing weather patterns. Hence a large optimization problem must be solved frequently, reducing the number of parameters allows for faster and more robust optimization. Second, by removing unimportant parameters the model is more robust and can be more easily integrated into a larger workflow, namely power scheduling. Third, we are able to learn more about the underlying system through these inferences, for instance, the unimportance of longitudinal information.

\section{Conclusion}\label{conclusion}
A new method for global sensitivity analysis of statistical model parameters was introduced. It addresses the challenges and exploits the problem structure specific to parameterized statistical models. The method is nearly fully automated; one step depends on the user's discretion, but this level of user specification is likely necessary for any alternative method. The proposed method also admits perturbation analysis at no additional computational cost, thus yielding sensitivities accompanied with certificates of confidence in them.

The method was motivated by, and applied to, a Gaussian process model aimed at wind simulation. Sensitivities were computed and the model parameterization simplified by removing 17\% of the model parameters. This simpler model was validated and shown to provide equal or superior predictive capability compared with the original model.

Our proposed method has two primary limitations. First, it relies heavily on Markov Chain Monte Carlo sampling for which convergence diagnostics are notoriously challenging.  Second, regularization may be needed to eliminate heavy-tailed distributions. Determining the regularization constant is simple in principle but may require drawing many samples to resolve. However, these limitations are classical and have been observed in various applications previously.

Global sensitivity analysis has seen much success analyzing parameter uncertainty for mathematical models. The framework presented in this paper provides the necessary tools to extend global sensitivity analysis to parameters of complex statistical models. To the authors knowledge, our approach is the first to systematically combined tools from mathematics and statistics to facilitate efficient global sensitivity analysis. There are a plurality of other approaches one can consider, which may yield similar or different results than our proposed method. Utilizing the loss function and its gradient is critical for our method's efficiency. Future work may include a broader exploration of how our approach compares with possible alternative methods, particularly those which have less dependence on the loss function. Analysis for groups of parameters may also be a useful extension of this work.

\appendix
\section*{Appendix}
\label{appendix}
This appendix contains the proofs for Proposition~\ref{Deltadiff}, Proposition~\ref{incfun}, Theorem~\ref{deltalimit}, Corollary~\ref{existence}, and Theorem~\ref{diffthm}.
\subsection*{Proof of Proposition~\ref{Deltadiff}}
Let $U(\delta)=e^{-\delta(\LL_\lambda(\theta)-\Lm)}$ and $V(\delta)=\int_B
e^{-\delta(\LL_\lambda(\theta)-\Lm)} \dd \theta$. Then
\begin{align*}
  \Delta(\delta)=\frac{1}{V(\delta)} \int_C U(\delta) \dd \theta \,.
\end{align*}
By using Theorem 6.28 in \cite{Klenke} with $(\LL_\lambda(\theta)-\Lm) e^{-\dm ( \LL_\lambda(\theta)-\Lm)}$ dominating $U'(\delta)$, we
have that $\int_C U(\delta) \dd \theta$ and $V$ are differentiable with
\begin{align*}
  \frac{d}{d\delta} \int_C U(\delta) \dd \theta&=-\int_C
  (\LL_\lambda(\theta)-\Lm)U(\delta)\dd \theta\,,\\
  V'(\delta)&=-\int_B (\LL_\lambda(\theta)-\Lm)e^{-\delta(\LL_\lambda(\theta)-\Lm)}
  \dd \theta\,.
\end{align*}
$\Delta$ is differentiable since $V(\delta)>0$, $\forall \delta>\dm$,
and applying the quotient rule for derivatives gives
\begin{align*}
  \Delta'(\delta)=\frac{\frac{\dd}{\dd \delta} \int_C
    U(\delta) \dd \theta}{V(\delta)}-\frac{V'(\delta)}{V(\delta)}
  \frac{\int_C U(\delta) \dd \theta}{V(\delta)}\,.
\end{align*}
Simple manipulations yields
\begin{align*}
  \Delta'(\delta)=(-1+\Delta(\delta)) \int_C (\LL_\lambda(\theta)-\Lm)
  \frac{U(\delta)}{V(\delta)} \dd \theta+\Delta(\delta) \int_{B \setminus
    C} (\LL_\lambda(\theta)-\Lm) \frac{U(\delta)}{V(\delta)} \dd \theta\,.
\end{align*}
Writing $\frac{U(\delta)}{V(\delta)}=p(\theta)$ and using the linearity of the integral completes the proof.\\
\qed

\subsection*{Proof of Proposition~\ref{incfun}}
From the proof of Proposition~\ref{Deltadiff} we have 
\begin{align*}
\Delta'(\delta)=(-1+\Delta(\delta)) \int_C (\LL_\lambda(\theta)-\Lm) \frac{U(\delta)}{V(\delta)} \dd \theta+\Delta(\delta) \int_{B \setminus C} (\LL_\lambda(\theta)-\Lm) \frac{U(\delta)}{V(\delta)} \dd \theta .
\end{align*}
Since 
\begin{align*}
&\theta \in C \implies \LL_\lambda(\theta) \le M_\lambda,\\
&\theta \in B \setminus C \implies \LL_\lambda(\theta) > M_\lambda,\\
&\Delta(\delta) \in [0,1],\\
&\int_{B \setminus C} p(\theta)\dd \theta=1-\Delta(\delta),
\end{align*}
we have
\begin{align*}
\Delta'(\delta)& > (-1+\Delta(\delta)) \int_C (M_\lambda-\Lm) \frac{U(\delta)}{V(\delta)} \dd \theta+\Delta(\delta) \int_{B \setminus C} (M_\lambda-\Lm) \frac{U(\delta)}{V(\delta)} \dd \theta\\
&=(-1+\Delta(\delta)) (M_\lambda-\Lm) \Delta(\delta)+\Delta(\delta)(M_\lambda-\Lm)(1-\Delta(\delta))\\
&=0.
\end{align*}
\qed

\subsection*{Proof of Theorem~\ref{deltalimit}}
Let $\alpha \in (0,1)$. Define
\begin{align*}
& \tilde{M}_\lambda=\frac{\LL(\theta')+M_\lambda}{2}<M_\lambda , \\
& \tilde{C}=\{ \theta \in B \vert \LL_\lambda(\theta) \le \tilde{M}_\lambda \}.
\end{align*}
By Assumption~\ref{a1}, $\LL_\lambda$ is continuous so $Vol(\tilde{C})>0$. Then $\exists \delta>\dm$ such that
\begin{align*}
\frac{Vol(B \setminus C)}{(1-\alpha) Vol(\tilde{C})} < e^{(M_\lambda-\tilde{M}_\lambda) \delta}.
\end{align*}
We want to show that
\begin{align*}
\frac{\int_C e^{-\delta(\LL_\lambda(\theta)-\Lm)} \dd \theta}{\int_B e^{-\delta(\LL_\lambda(\theta)-\Lm)}\dd \theta}>\alpha \iff (1-\alpha) \int_C e^{-\delta(\LL_\lambda(\theta)-\Lm)} \dd \theta > \alpha \int_{B \setminus C} e^{-\delta(\LL_\lambda(\theta)-\Lm)} \dd \theta .
\end{align*}
It is enough to show
\begin{align*}
\int_{B \setminus C} e^{-\delta(\LL_\lambda(\theta)-\Lm)} \dd \theta < (1-\alpha) \int_C e^{-\delta(\LL_\lambda(\theta)-\Lm)} \dd \theta .
\end{align*}
If $\theta \in B \setminus C$, then
\begin{align*}
\LL_\lambda(\theta)>M_\lambda \implies e^{-\delta(\LL_\lambda(\theta)-\Lm)} \le e^{-\delta(M_\lambda-\Lm)}; 
\end{align*}
hence it is enough to show that
\begin{align*}
e^{-\delta(M_\lambda-\Lm)} Vol(B \setminus C) < (1-\alpha) \int_C e^{-\delta(\LL_\lambda(\theta)-\Lm)} \dd \theta .
\end{align*}
Since the exponential is nonnegative and $\alpha \in (0,1)$, it is equivalent to show
\begin{align*}
\frac{Vol(B \setminus C)}{1-\alpha} < \int_{\tilde{C}} e^{-\delta(\LL_\lambda(\theta)-M_\lambda)} \dd \theta .
\end{align*}
If $\theta \in \tilde{C}$, then
\begin{align*}
\LL_\lambda(\theta) \le \tilde{M}_\lambda \implies \int_{\tilde{C}} e^{-\delta(\LL_\lambda(\theta)-M_\lambda)} \dd \theta \ge \int_{\tilde{C}} e^{-\delta(\tilde{M}_\lambda-M_\lambda)} \theta=e^{-\delta(\tilde{M}_\lambda-M_\lambda)} Vol(\tilde{C}) .
\end{align*}
But
\begin{align*}
\frac{Vol(B \setminus C)}{(1-\alpha) Vol(\tilde{C})} < e^{(M_\lambda-\tilde{M}_\lambda) \delta}
\end{align*}
by our construction of $\delta$.\\
\qed

\subsection*{Proof of Corollary~\ref{existence}}
Since $B$ is bounded, $e^{-\delta \LL_\lambda(\theta)} \in L^1(B)$ $\forall \delta \in \R$. Mimicking the argument of Proposition~\ref{Deltadiff}, $\Delta$ is differentiable and hence continuous at $\dm$. Since $\alpha \in (0,1)$, Theorem~\ref{deltalimit} gives $\exists \delta>\dm$ such that $\Delta(\dm)<\alpha<\Delta(\delta)$. Then existence holds by the intermediate value theorem. Uniqueness follows from Proposition~\ref{incfun}.\\
\qed

\subsection*{Proof of Theorem~\ref{diffthm}}
Let  
\begin{align*}
U(\delta)=\int_B \vert \theta_k \vert e^{-\delta(\LL_\lambda(\theta)-\Lm)}\dd \theta 
\end{align*}
\begin{align*}
V(\delta)=\int_B \left\vert \frac{\partial \LL}{\partial \theta_k}(\theta) \right\vert e^{-\delta(\LL_\lambda(\theta)-\Lm)}\dd \theta 
\end{align*}
and
\begin{align*}
W(\delta)=\int_B e^{-\delta(\LL_\lambda(\theta)-\Lm)} \dd \theta .
\end{align*}
Then we have
\begin{align*}
F_k(\delta)=\frac{U(\delta)}{W(\delta)}\frac{V(\delta)}{W(\delta)} .
\end{align*}
Note that $W(\delta)>0$ $\forall \delta>0$, so it is enough to show that $U$, $V$, and $W$ are differentiable. Theorem 6.28 in \cite{Klenke} gives the result using
\begin{align*}
\left\vert \theta_k \right\vert (\LL_\lambda(\theta)-\Lm) e^{-\dm (\LL_\lambda(\theta)-\Lm)},
\end{align*}
\begin{align*}
\left\vert \frac{\partial \LL}{\partial \theta_k}(\theta) \right\vert (\LL_\lambda(\theta)-\Lm) e^{-\dm (\LL_\lambda(\theta)-\Lm)},
\end{align*}
and
\begin{align*}
(\LL_\lambda(\theta)-\Lm) e^{-\dm (\LL_\lambda(\theta)-\Lm)}
\end{align*}
to dominate the derivatives of the integrands of $U$, $V$, and $W$, respectively. Applying Theorem 6.28 in \cite{Klenke} to $U(\delta)$, $V(\delta)$, and $W(\delta)$ yields
\begin{align*}
U'(\delta)=-\int_B \left\vert \theta_k \right\vert (\LL_\lambda(\theta)-\Lm) e^{-\delta(\LL_\lambda(\theta)-\Lm)} \dd \theta
\end{align*}
\begin{align*}
V'(\delta)=-\int_B \left\vert \frac{\partial \LL}{\partial \theta_k}(\theta) \right\vert (\LL_\lambda(\theta)-\Lm) e^{-\delta(\LL_\lambda(\theta)-\Lm)} \dd \theta
\end{align*}
and 
\begin{align*}
W'(\delta)=-\int_B (\LL_\lambda(\theta)-\Lm)e^{-\delta (\LL_\lambda(\theta)-\Lm)}\dd \theta .
\end{align*}
An application of the product and quotient rules to $F_k$ yields 
\begin{align*}
F_k'(\delta)=\left( \frac{U'(\delta)}{W(\delta)}-\frac{W'(\delta)}{W(\delta)} \frac{U(\delta)}{W(\delta)} \right) \frac{V(\delta)}{W(\delta)} + \left( \frac{V'(\delta)}{W(\delta)}-\frac{W'(\delta)}{W(\delta)} \frac{V(\delta)}{W(\delta)} \right) \frac{U(\delta)}{W(\delta)} .
\end{align*}
Basic algebra along with the fact that $\frac{e^{-\delta (\LL_\lambda(\theta)-\Lm)}}{W(\delta)}=p(\theta)$ yields the result.\\
\qed

\bibliographystyle{siamplain}
\bibliography{Global_sensitivity_analysis_for_statistical_model_parameters_biblio}

%\iffalse
\vfill
\begin{flushright}
  \scriptsize \framebox{\parbox{3.2in}{
The submitted manuscript has been created by UChicago Argonne, LLC, Operator of Argonne National Laboratory (“Argonne”). Argonne, a U.S. Department of Energy Office of Science laboratory, is operated under Contract No. DE-AC02-06CH11357. The U.S. Government retains for itself, and others acting on its behalf, a paid-up nonexclusive, irrevocable worldwide license in said article to reproduce, prepare derivative works, distribute copies to the public, and perform publicly and display publicly, by or on behalf of the Government.  The Department of Energy will provide public access to these results of federally sponsored research in accordance with the DOE Public Access Plan. \url{http://energy.gov/downloads/doe-public-access-plan}.}}
\normalsize 
\end{flushright}
%\fi

\end{document}